\documentclass[aps,prl,twocolumn,superscriptaddress,amsmath,amssymb]{revtex4-1}
\usepackage{mathrsfs}
\usepackage{graphicx}
\usepackage{dcolumn}
\usepackage{bm}
\usepackage{amsmath}
\usepackage{amsfonts}

\begin{document}

\title{Spin Liquid Condensate of Spinful Bosons}
\author{Biao Lian}
\affiliation{Department of Physics, McCullough Building, Stanford University, Stanford, California 94305-4045, USA}
\author{Shoucheng Zhang}
\affiliation{Department of Physics, McCullough Building, Stanford University, Stanford, California 94305-4045, USA}
\date{\today}

\begin{abstract}
We introduce the concept of a bosonic spin liquid condensate (SLC), where spinful bosons in a lattice form a zero-temperature spin disordered charge condensate that preserves the spin rotation symmetry, but breaks the U($1$) symmetry due to a spinless order parameter with charge one. It has an energy gap to all the spin excitations. We show that such SLC states can be realized in a system of spin $S\ge2$ bosons. In particular, we analyze the SLC phase diagram in the spin $2$ case using a mean-field variational wave function method. We show there is a direct analogy between the SLC and the resonating-valence-bond (RVB) state.
\end{abstract}
\maketitle

Pursuing new states of matter is always one of the principal aims of condensed matter physics. Compared to fermions which have various novel phases, bosons appear to be somewhat trivial. The zero temperature phases of bosons traditionally fall into three classes: the Mott insulator (MI), the Bose glass, and the Bose-Einstein condensate (BEC) on a single-particle state \cite{Fisher1989}. For example, the spinful bosons in ultra-cold atom systems \cite{Stenger1998} usually condense onto a particular spinor state, forming a spinor BEC that spontaneously breaks both the U($1$) and the spin rotation SU($2$) symmetries \cite{Ho1998,Ohmi1998,Ciobanu2000,Ueda2002,Barnett2006,Santos2006,Diener2006,Kawaguchi2011,Lian2012}. Whether bosons can form other zero temperature phases such as a condensate not on a single particle state (non-SPS), is an interesting question that has been discussed extensively\cite{Mueller2006}. In particular, a condensate of spinful bosons is a non-SPS condensate if it breaks only the U($1$) symmetry but preserves the spin rotation SU($2$) symmetry \cite{Mueller2006}. So far, however, the attempts along this direction never come with a rigorous example. Early proposed states like the spin-paired condensate of spin $1$ bosons \cite{Law1998} are shown to be unstable ``Shr{\"o}dinger cat" states with no spin gap in the thermodynamic limit \cite{Ho2000b,Mueller2006}, though they may be favored at a finite temperature \cite{Mukerjee2006,Natu2011}. Later works also discussed the possibility of stabilizing such condensates with interactions and lattices, but gave no definite answer \cite{Demler2002,Zhou2003,Ruostekoski2007}. Looking for non-SPS condensates with spin rotational invariance is nonetheless interesting, and is in some sense reminiscent of realizing spin liquids in fermionic systems. Such condensates may give a lot of nontrivial physics such as the spin-charge separation, in analogy to those in spin liquids \cite{Balents2010}.

%Bose-Einstein condensate (BEC) is known as a bosonic many-body state characterized by a spontaneously broken global U($1$) symmetry. With the development of cold atom techniques in the past decades \cite{Stenger1998}, the BEC of spinful bosons has attracted a great deal of attention. Compared to scalar bosons which only have a U($1$) symmetry, spinful bosons possess an additional SU($2$) spin rotational symmetry. Usually, spinful bosons at low temperatures form a spinor BEC, which has a spinor order parameter that breaks both the U($1$) and SU($2$) symmetries \cite{Ho1998,Ohmi1998,Ciobanu2000,Ueda2002,Barnett2006,Santos2006,Diener2006,Kawaguchi2011,Lian2012}. A natural question is whether a spin rotationally invariant BEC can exist. Law, Pu and Bigelow \cite{Law1998} first proposed a spin-paired BEC ground state for spin $1$ bosons that preserves the SU($2$) symmetry, but the state was shown to be an unstable ``Shr{\"o}dinger cat" state and gives way to the spinor BEC polar phase in the thermodynamic limit \cite{Ho2000b,Mueller2006}. Later attempts along this direction are mostly focused on spin $1$ bosons\cite{Demler2002,Zhou2003}, without rigorous evidences. Other studies have proposed that a spinor BEC may turn into a spin rotationally invariant condensate at finite temperatures due to thermal fluctuations \cite{Mukerjee2006,Natu2011}. However, a spin rotationally invariant BEC at zero temperature has never been rigorously found in the previous studies. Looking for such a condensate is in some sense reminiscent of searching for spin liquids in fermionic systems.

In this letter, we introduce the concept of a bosonic spin liquid condensate (SLC), which is a robust non-SPS spin disordered charge condensate at zero temperature. It is defined as a state of spinful bosons that preserves the spin rotation symmetry with a spin gap, but has a spontaneously broken U($1$) symmetry due to a locally defined spinless order parameter that carries a U($1$) charge.
%due to an off diagonal long range order (ODLRO) in a $k$-particle density matrix ($k>1$) \cite{Yang1962,supplement}.
We show a way to construct an SLC for spin $S\ge2$ bosons in a lattice. The SLC has a direct analogy to the resonating valence bond (RVB) state for cuprate superconductors proposed by Anderson \cite{Anderson1987,Kivelson1987}, where bosons are free to move with their spins confined in short range RVBs. We calculate explicitly the SLC phase diagram for spin $2$ via a mean-field variational wave function method, and study the spin excitations and the Goldstone mode of the state. A lot more physics in SLC is awaiting exploration.

%In this letter, we introduce the concept of a bosonic spin liquid condensate (SLC), which is a robust spin disordered charge condensate at zero temperature. It possesses charge superfluidity, with an energy gap to all spin excitations. We show that SLC phases generally exist for spin $S\ge2$ bosons in a lattice. Bosons in the condensate move around, accompanied by short range singlet resonating valence bonds (RVBs), which is in direct analogy to the electrons in an RVB state for high temperature cuprates proposed by Anderson\cite{Anderson1987,Kivelson1987}. The letter is organized as follows. First, we sketch the idea and give the definition of SLC, using spin $2$ bosons  in a lattice as a example. Next, we calculate the phase diagram for spin $2$ bosons via a variational wave function method, and verify the existence of SLC. We then show the analogy between SLC and RVB, and study the spin excitations and the Goldstone mode in SLC. Lastly, we briefly make a connection between our theory and the experiments.

%Bosons in a translationally invariant lattice fall into either a Mott insulator or a BEC superfluid at zero temperature \cite{Fisher1989}.
In a Mott insulator where interaction dominates, every site is in the lowest on-site state with a definite particle number $n$. As the hopping between sites increases, a coherent superposition of several on-site states with different particle numbers is preferred to gain the kinetic energy, which is the BEC state. For spin $S$ bosons, the coherent superposition usually induces a non-vanishing spinor order parameter $\varphi_m=\langle \psi_{i,m}\rangle$, where $\psi_{i,m}$ is the boson field operator, and one obtains a spinor BEC breaking the spin rotation symmetry. However, suppose the lowest two on-site states are spin singlet states close to each other in energy, while the energies of all the other states are much higher. When the hopping is not too large, the system may prefer a superposition of the two singlet states only, forming a spin singlet condensate with $\langle \psi_{i,m}\rangle=0$ ensured by the spin rotation symmetry. It is nonetheless possible to construct a non-vanishing order parameter that carries a U($1$) charge and total spin zero (see Eq. (\ref{order_parameter})).

To find such on-site state spectrums explicitly, we examine the spin Bose Hubbard model widely used for describing spin $S$ bosons in a lattice. The Hamiltonian $H=H_I+H_t$ can be written in the following two parts:
\begin{equation}\label{H}
\begin{split}
&H_I=-\mu\sum_{i} \hat{n}_i+\frac{1}{2}\sum_i\left[\sum_{J=0}^{S}U_{2J}\hat{\mathcal{P}}^S_{2J}(i)\right],\\
&H_t=-\sum_{\langle ij\rangle,m}\left(t\psi_{i,m}^\dag\psi_{j,m}+h.c.\right) \ ,
\end{split}
\end{equation}
where $\mu$ is the chemical potential, $U_{2J}\ge0$ is the on-site Hubbard interaction energy between two bosons of total spin $2J$, and $t$ is the nearest hopping amplitude. $\psi_{i,m}$ is the boson field operator, where $i$ and $j$ label the lattice sites, $m$ denotes the spin $z$ component. $\hat{n}_i=\sum_m\psi_{i,m}^\dag\psi_{i,m}$ is the particle number on site $i$. The non-negative projection operator $\hat{\mathcal{P}}^S_{2J}(i)$ is defined as $\hat{\mathcal{P}}^S_{2J}(i)=\sum_m\mathcal{A}^\dag_{2Jm}(i)\mathcal{A}_{2Jm}(i)$ with $\mathcal{A}_{2Jm}(i)=\sum_{m'}\langle 2J,m|S,m';S,m-m'\rangle\psi_{i,m'}\psi_{i,m-m'}$, where $\langle 2J,m|S,m';S,m-m'\rangle$ is the Clebsch-Gordan coefficient. The on-site state energy spectrum can be obtained by diagonalizing $H_I$. For spin $S=1$ bosons, we find it impossible to have both the lowest two on-site states be spin singlets (see supplementary material \cite{supplement}). However, this is possible for bosons with spin $S\ge2$.

\begin{figure}
\includegraphics[width=3.2in]{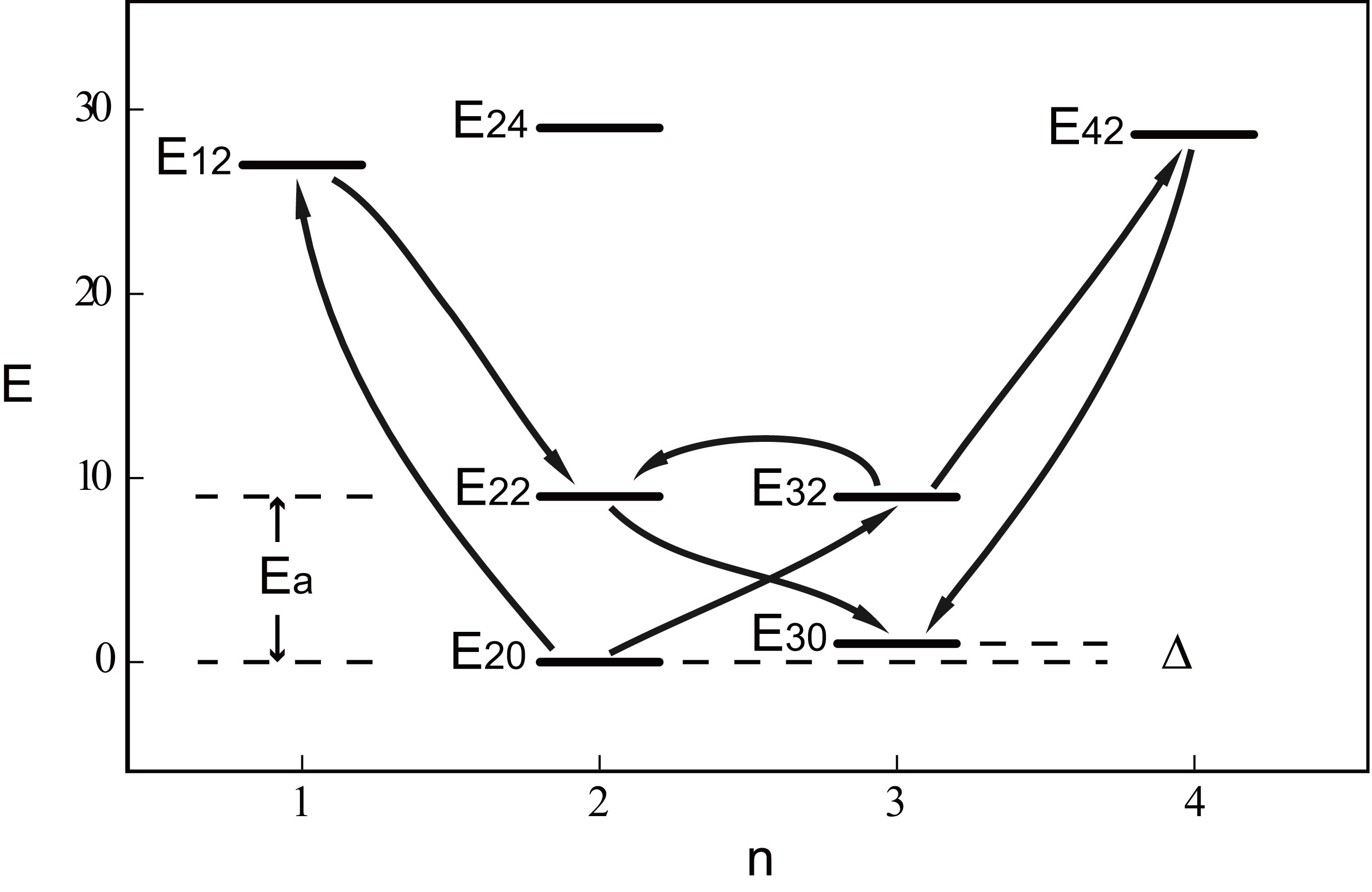}
\caption{The on-site energy levels computed for $U_4=3U_2=30U_0$ and $\mu=\mu_0-\Delta$, where $\mu_0=3U_2-U_0$. The $x$ and $y$ axes are the particle number $n$ and the energy $E$ in units of $U_0$ respectively. $E_{nl}$ represents the energy of the on-site state with particle number $n$ and total spin $l$. The arrows show the transition process from $|2,0,0\rangle_i$ to $|3,0,0\rangle_i$ via three time hoppings, which must overcome an activation energy $E_a$. \label{hop}}
\end{figure}

We study spin $2$ bosons in this letter to show how an SLC can be realized. First, we need to find a parameter regime where the lowest two on-site states are singlet states. Generally, the on-site states can be labeled as $|n,l,m,\gamma\rangle_i$, where $n$ is the particle number, $l$ is the total spin of the $n$ particles, $m$ is the $z$-component of the total spin, and $\gamma$ is an additional quantum number \cite{Snoek2009,supplement}. In our discussion, we shall omit the label $\gamma$, since all the states involved are distinguishable through their $n,l,m$ labels \cite{supplement}. The minimal two singlet states of spin $2$ bosons are the dimer state $|2,0,0\rangle_i$ and the trimer state $|3,0,0\rangle_i$, consisting of two and three bosons respectively \cite{Zhou2006,Lian2013}. We find when $U_0<U_2<(36U_4+49U_0)/85$ and $\mu=\mu_0=3U_2-U_0$, the two singlet states become degenerate and have the lowest on-site energy. To see this explicitly, we plot the on-site state energy spectrum for $U_4=3U_2=30U_0$ and $\mu=\mu_0-\Delta$ in Fig. \ref{hop}. The on-site energy of each state $|n,l,m\rangle_i$ is denoted by $E_{nl}$. An easy calculation shows that $E_{30}-E_{20}=\Delta$. For later convenience, we define $E_a$ as the energy difference between the third and the lowest energy levels, as is shown in Fig. \ref{hop}. The desired regime is then $|\Delta|\ll E_a$.

Then, for the SLC ground state to be favored, the transition amplitude between the two on-site states $|2,0,0\rangle_i$ and $|3,0,0\rangle_i$ must be large enough. The transition amplitude originates from the hopping energy $H_t$. However, the tunneling from $|2,0,0\rangle_i$ to $|3,0,0\rangle_i$ cannot be achieved by a single hop, since a single hop will change a singlet state to a state with total spin $l=2$. For the transition to occur, the site must hop with nearby sites for at least $3$ times, during which the site must go to two intermediate states with energy of order $E_a$, as is shown in Fig. \ref{hop}. In perturbation theory, this mechanism gives us a effective hopping amplitude between the singlet dimer state and the trimer state $t_{eff}=t^3/E_a^2$. When the hopping amplitude increases to $|t_{eff}/\Delta|\sim1$, the system would prefer a coherent superposition of the dimer and trimer states. On the other hand, the condition $|\Delta|\ll E_a$ ensures that $|t/E_a|\ll1$, so the superposition with any other states is not yet favorable, and the system will form an SLC. We can write down a spin-rotationally invariant operator $\hat{\lambda}_i$ \cite{supplement}:
%order parameter for the SLC showing the U($1$) symmetry breaking  \cite{supplement}:
\begin{equation}
\hat{\lambda}_i=\sum_{m,m'=-2}^{2}D_{m'm}\psi_{i,m'}\psi_{i,m}\psi^\dag_{i,m+m'}
\label{order_parameter}
\end{equation}
where the coefficient $D_{m'm}=(-1)^{m+m'}\langle 0,0|2,-m-m';2,m+m'\rangle\langle2,m+m'|2,m';2,m\rangle$, and define an order parameter $\lambda=\langle \hat{\lambda}_i\rangle$. It can be shown that
\[[\hat{\mathbf{S}}_i,\hat{\lambda}_i]=0\ ,\qquad [\hat{n}_i,\hat{\lambda}_i]=-\hat{\lambda}_i\ ,\]
where $\hat{\mathbf{S}}_i$ is the total spin of site $i$, hence $\hat{\lambda}_i$ is a charge $1$ operator. The minimal flux of a superfluid vortex is thus $2\pi$. The definition of $\hat{\lambda}_i$ again indicates the significance of hopping $3$ times in the SLC. The existence of the order parameter $\lambda\propto\langle\psi_i\psi_i\psi^\dag_i\rangle$ also shows that there is an off diagonal long range order (ODLRO) in the $3$-particle density matrix $\rho^{(3)}$. The leading eigenvalue of $\rho^{(3)}$ can be estimated as $r_3\sim\sum_{j}\langle\hat{\lambda}^\dag_i\hat{\lambda}_j\rangle\approx N_S|\lambda|^2\sim \mathcal{O}(N)$, where $N_S$ and $N$ are the total number of sites and bosons respectively. However, there is no ODLRO in $\rho^{(1)}$ or $\rho^{(2)}$ \cite{Yang1962,supplement}. This implies that any boson participating in this condensate is ``dressed": Its spin is fully screened by a local virtual particle-hole pair, while its charge remains unchanged.
%One can think of this process as a strong ``spin vacuum polarization" \cite{vp}.
%In fact, it is a generally true that the ODLRO of an SLC arises in a $j$-particle density matrix $\rho^{(j)}$ ($j>1$) instead of $\rho^{(1)}$ \cite{supplement}.
This is a key difference between SLC and the spin-paired charge $2$ condensate in Ref. \cite{Law1998}, where the ODLRO arises in $\rho^{(1)}$ \cite{Mueller2006}.

%Under a U($1$) transformation $\lambda$ transforms as $\lambda\rightarrow e^{i\phi}\lambda$ like a charge $1$ boson operator, implying the state is a condensate of charges with elementary charge $1$.
%This is a prominent difference between SLC and the spin-paired BEC with elementary charge $2$ \cite{Law1998}.

To confirm the existence of the SLC phase, we have proposed a mean-field variational wave function for the ground state of the system:
\begin{widetext}
\begin{equation}\label{wavefunction}
%\begin{split}
|SLC\rangle=Sym\prod_{\langle ij\rangle}\Big[u+\sum_m\left(v\psi_{i,m}^\dag\psi_{j,m}+h.c.\right)\Big] \times\prod_i\Big(\alpha|2,0,0\rangle_i+\beta|3,0,0\rangle_i\Big)\ ,
%&\qquad\qquad\qquad\qquad\qquad
%\end{split}
\end{equation}
\end{widetext}
in terms of four variational parameters $u$, $v$ and $\alpha$, $\beta$ satisfying $|\alpha|^2+|\beta|^2=1$. The notation $Sym$ represents a symmetrization of all nearest-site bonds $\langle ij\rangle$ so that there is no preferred sequence of $\langle ij\rangle$ in the product. The product operator in the front defined on $\langle ij\rangle$ represents the quantum fluctuation induced by $H_t$, and establishes the correlation between lattice sites. In the limit $|t/E_a|\ll 1$, a simple estimation gives $v/u\sim t/E_a$, and the quantum fluctuation is weak. However, it is indispensable in the calculation of the energy contribution of the $3$ times hoppings. It is easy to show the order parameter of SLC defined in Eq. (\ref{order_parameter}) is given by $\lambda=\sqrt{\frac{12}{5}}\left[1+\mathcal{O}(\left|\frac{v}{u}\right|^2)\right]\alpha^*\beta$. All we need to do then is to minimize the variational energy per site $\mathcal{E}_G=\langle H\rangle/N_S$, and the SLC state is characterized by $\lambda\propto\alpha^*\beta\neq0$.

\begin{figure}
\includegraphics[width=3.2in]{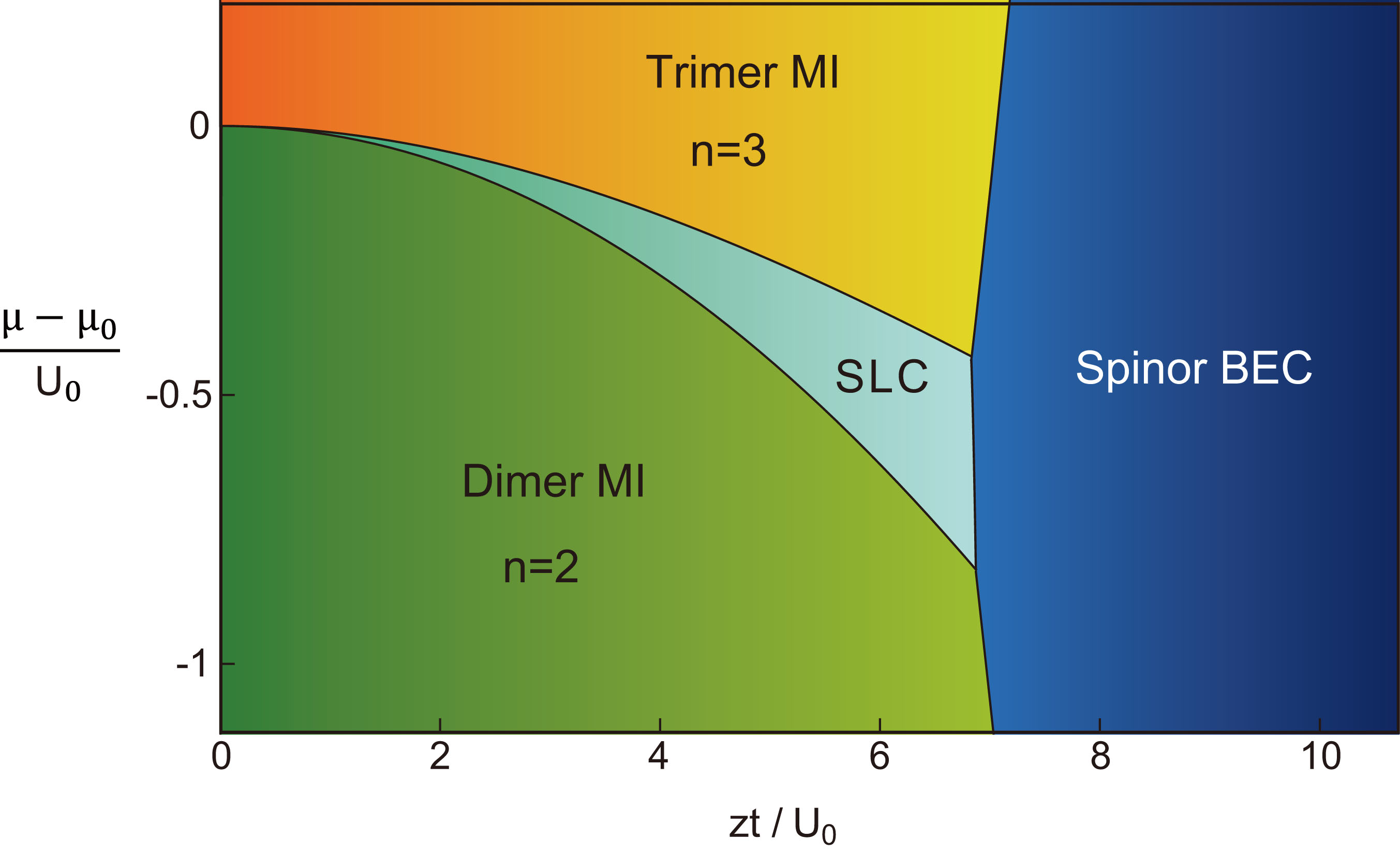}
\caption{The phase diagram for $U_4=3U_2=30U_0$ and $z=4$ with respect to the chemical potential $\mu$ and the hopping $zt$. $E_a\approx10U_0$ as is shown in Fig. \ref{hop}. Dimer MI and trimer MI stands for the singlet Mott insulators with $n=2$ and $n=3$ respectively. In the large $t$ limit the system becomes a spinor BEC (nematic phase according to Ref. \cite{Ciobanu2000,Ueda2002,Barnett2006}). \label{phasediagram}}
\end{figure}

Since $v/u$ is a small parameter, as a good approximation we can derive the variational energy per site $\mathcal{E}_G$ up to the quadratic order $|v/u|^2$. We choose $t$ real and positive so that there is no flux in the lattice, which allows $\alpha$, $\beta$ to be set real. Up to the quadratic order, the normalization condition for the wave function is given by $|u|^2+2z(2+\beta^2)(7+\beta^2)|v|^2/5=1$, where $z$ is the number of nearest neighbours of a site. After some calculations, the variational energy $\mathcal{E}_G$ can be expressed as \cite{supplement}:
\begin{equation}
\begin{split}
&\mathcal{E}_G=|u|^2\beta^2\Delta-\frac{2}{5}zt\text{Re}(u)\text{Re}(v)\left(2+\beta^2\right)\left(7+\beta^2\right)\\
&\quad-\frac{36}{5}zt|v|^2\left(1-\beta^2\right)\beta^2+\frac{1}{2}z|v|^2V\left(\beta^2\right),
\end{split}
\end{equation}
where $V(\beta^2)$ is a quadratic function of $\beta^2$ given in the supplementary material \cite{supplement}, which is of order $E_a$. In particular, we minimize the variational energy $\mathcal{E}_G$ for $U_4=3U_2=30U_0$ and $z=4$, and the phase diagram is shown in Fig. \ref{phasediagram}. The SLC phase arises in the regime we expected. The boundary of spinor BEC phase is obtained separately by the known Gutzwiller method \cite{supplement}. In the limit $t/E_a\ll1$, the phase boundaries between SLC and the Mott insulators take the following form:
\begin{equation}
\Delta+a_{\pm}\frac{zt^2}{E_a}=\pm b_\pm^2\frac{zt^3}{E_a^2}\ ,
\end{equation}
where $a_\pm$ and $b_\pm$ are dimensionless factors depending on the interaction parameters $U_{2J}$ only. This result agrees with our expectation $|t_{eff}/\widetilde{\Delta}|\sim1$ for the phase transition, except for that the on-site energy difference $\widetilde{\Delta}=\Delta+a_{\pm}zt^2/E_a$ is corrected by a second order virtual hopping perturbation. The SLC finally becomes unstable against quantum fluctuations when $t/E_a\sim1$, and the spinor BEC phase takes charge.

\begin{figure}
\includegraphics[width=3.2in]{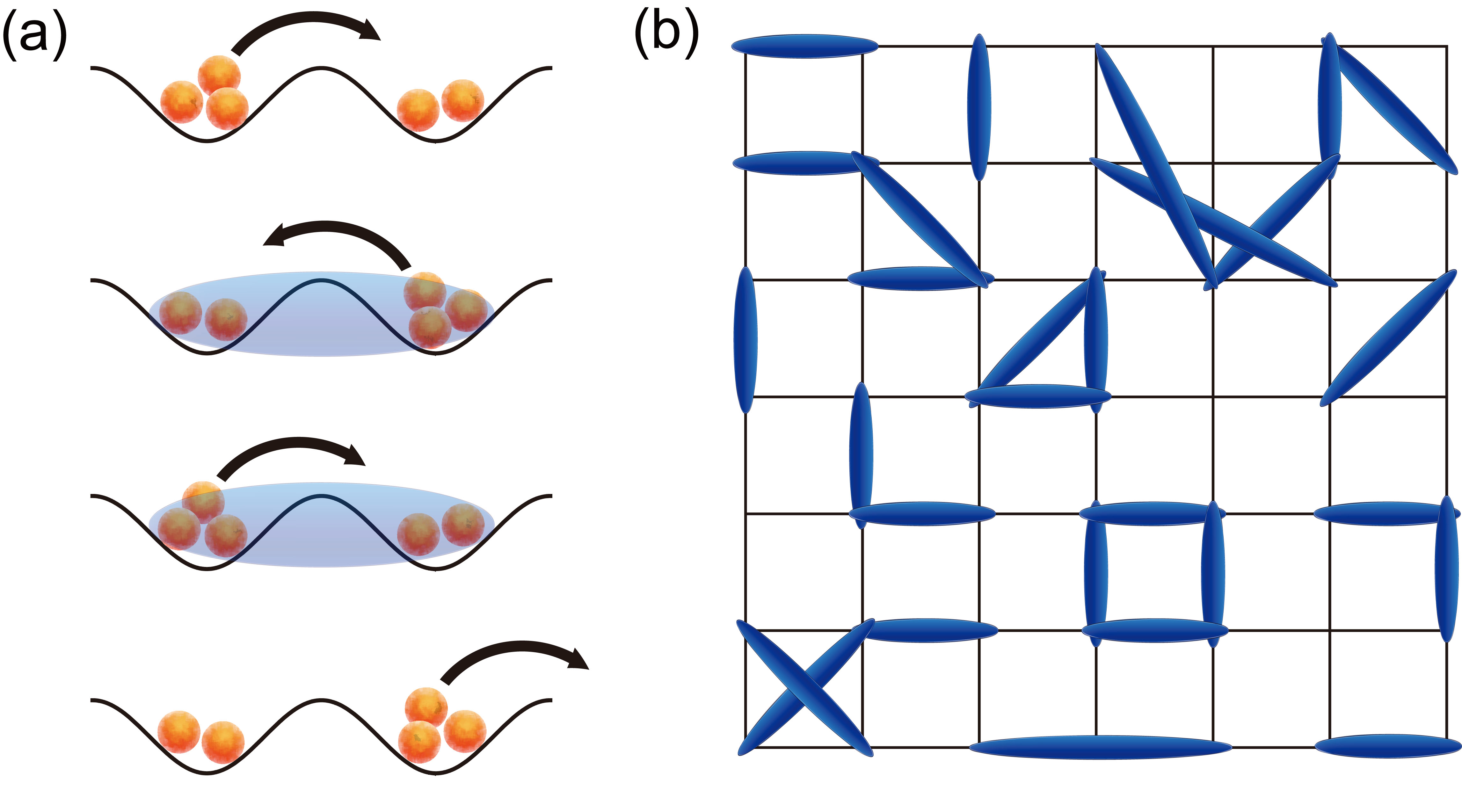}
\caption{(a). How a particle moves in SLC. Suppose initially two nearby sites are in states $|3,0,0\rangle_i$ (left) and $|2,0,0\rangle_i$ (right) respectively. Hopping once induces a singlet valence bond between the two sites and costs energy of order $E_a$. By two more hoppings, the two sites become spin singlets again and the valence bond is erased, while a particle moves to the right site. (b) Illustration of various resonating valence bonds created and annihilated due to the motion of particles in SLC. \label{RVB}}
\end{figure}

%It satisfies the identity $\sum_J\hat{\mathcal{P}}^S_J(i)=\hat{n}_i(\hat{n}_i-1)$.

%The Hamiltonian for spin $S$ bosons in a lattice with an SU($2$) spin rotational symmetry.

We now take a closer look at the SLC state, and show it has an energy gap to all the spin excitations. To possess the charge superfluidity, particles in the SLC have to move around and undergo macroscopic ring exchanges \cite{Feynman1953}. However, the motion of the particles must not break the spin rotational symmetry. This is in fact naturally achieved by hopping $3$ times. As is shown in Fig. \ref{RVB}(a), suppose a particle in a trimer state is to move to a nearby site dimer state on the right. By hopping once, both sites become high energy states of total spin $l=2$. Due to the spin rotational symmetry of $H_t$, they form a spin singlet valence bond of length $1$. To lower the energy, the particle must hop back to the left site and then forward to the right site, so that the left and right sites become singlet states again, with the particle numbers exchanged. By such a process, the particles are free to move in the lattice, with their spins confined in singlet valence bonds created and erased resonantly. SLC is in this sense analogous to the dopped RVB state suggested by Anderson for high temperature cuprates \cite{Anderson1973,Anderson1987,Kivelson1987}, where electrons form resonating valence bonds while breaking the U($1$) symmetry. In principle, singlet valence bonds of length $L>1$ can arise by hopping more times, as is shown in Fig. \ref{RVB}(b). According to Eq. (\ref{wavefunction}), the amplitude of creating a singlet valence bond of length $L$ is approximately $(v/u)^L\propto(t/E_a)^L$, decaying exponentially as a function of $L$. The singlet valence bonds in SLC are thus short range, indicating that all the spin excitations are gapped. The spin correlation length is roughly $\xi_S\approx1/\ln (y_cE_a/t)$, where $y_c=(t/E_a)_c$ is the critical value for the phase transition from SLC to a spinor BEC. More physics in the RVB studies may come in parallel in SLC, such as the mapping to a loop gas model \cite{Sutherland1988}. The loop gas model for SLC has an action similar to that of the loop gas model for RVB, except for that it does not require all the sites to be covered by loops \cite{supplement}.

By the Goldstone theorem, the SLC should have a gapless mode that is spinless, corresponding to the broken U($1$) symmetry. This mode can be derived at low energies by perturbation of the wave function $|SLC\rangle$. For convenience, we rewrite $\psi_{i,m}=\sqrt{\hat{n}_{i,m}}e^{i\phi_{i,m}}$, where $n_{i,m}$ and $\phi_{i,m}$ satisfies $[\hat{n}_{i,m},\phi_{j,m'}]=i\delta_{ij}\delta_{mm'}$. The global phase can then be expressed as $\phi_i=\sum_m\phi_{i,m}/5$, which is rotationally invariant and satisfies the commutation relation $[n_i,\phi_j]=i\delta_{ij}$, where $\hat{n}_i$ is the particle number. We denote the fluctuations of particle number and global phase by $\delta \hat{n}_i=\hat{n}_i-\langle\hat{n}_i\rangle$ and $\delta\phi_i=\phi_i-\langle \phi_i\rangle$. With $\alpha$ and $\beta$ real and positive, we have $\langle\hat{n}_i\rangle=2+\beta^2$ and $\langle \phi_i\rangle=0$. By adding the two fluctuations $\delta\hat{n}_i$ and $\delta\phi_i$ to the variational wave function in Eq. (\ref{wavefunction}), one obtains a low energy effective Hamiltonian:
\begin{equation}
\begin{split}
&H_{eff}=\sum_{\mathbf{k}}\Big[\frac{1}{8\beta^2}\frac{\mbox{d}^2\mathcal{E}_G}{\mbox{d}\beta^2}\delta\hat{n}_\mathbf{k}\delta\hat{n}_\mathbf{-k} \\ &\qquad\qquad\qquad+\frac{18}{5}\frac{zt|v|^2(1-\beta^2)\beta^2}{d}\mathbf{k}^2\delta\phi_\mathbf{k}\delta\phi_\mathbf{-k}\Big] ,
\end{split}
\end{equation}
where $d$ is the dimension of the system. This is a Hamiltonian of harmonic oscillators that can be easily diagonalized, and hence we get a linear energy dispersion $\omega_\mathbf{k}=v_sk$, where $v_s=\sqrt{9zt|v|^2(1-\beta^2)(\mbox{d}^2\mathcal{E}_G/\mbox{d}\beta^2)/5d}$ is the superfluid velocity. In the limit $t/E_a\ll1$, the velocity is asymptotically $v_s\propto \alpha\beta t^{5/2}/E_a^{3/2}$. This linear Goldstone mode is obviously spinless, and thus describes the charge fluctuations only. The distinct energy dispersions of spin and charge excitations naturally lead to a spin-charge separation in SLC.

Using the wave function $|SLC\rangle$, one can easily show that the $1$-particle density matrix $\rho^{(1)}$ has the form
\begin{equation}
\rho^{(1)}_{mm'}(\mathbf{x}_i,\mathbf{x}_j)=\langle\psi_{i,m}^\dag\psi_{j,m'}\rangle\approx\delta_{mm'}e^{-|\mathbf{x}_i-\mathbf{x}_j|/\xi_S}\ ,
\end{equation}
where $\xi_S$ is the spin correlation length defined above. This demonstrates that $\rho^{(1)}$ has no eigenvalue of order $N$. In contrast, in the limit $|\mathbf{x}_{i}-\mathbf{x}_{j}|\rightarrow\infty$,
\begin{equation}
\begin{split}
&\rho^{(3)}_{m_1m_2m_3m_4m_5m_6}(\mathbf{x}_{i},\mathbf{x}_{i},\mathbf{x}_{j},\mathbf{x}_{i},\mathbf{x}_{j},\mathbf{x}_{j})\\
=&\delta_{m_1+m_2-m_4}\delta_{m_5+m_6-m_3}D_{m_1m_2}^*D_{m_5m_6}|\lambda|^2\neq0\ ,
\end{split}
\end{equation}
which indicates $\rho^{(3)}$ has an eigenvalue of order $N$ \cite{supplement}.

%   &=\langle \psi^\dag_{i,m_1}\psi^\dag_{i,m_2}\psi^\dag_{j,m_3}\psi_{i,m_4}\psi_{j,m_5}\psi_{j,m_6}\rangle\\
%Similarly, one may show there is also no eigenvalue of order $N$ in the $2$-particle density matrix $\rho^{(2)}$.

Though we have only analyzed the SLC for spin $2$ bosons, this way of constructing SLC phases is quite general. For bosons with spin $S>2$ in a lattice, there are more Hubbard parameters $U_{2J}$, and more low energy singlet on-site states can be constructed \cite{Lian2013}. We therefore expect more SLC phases to exist in higher spin boson systems.

Finally, we briefly discuss on the experimental realization and observation of SLC in cold atom experiments. The Bose-Hubbard model can be implemented by trapping bosonic atoms into an optical lattice. The Mott-superfluid physics of spinless bosons has been observed in several experiments \cite{Greiner2002,Gerbier2005,Stoferle2004,Spielman2007,Sherson2010,Bakr2010}. For hyperfine spin $2$ atoms $^{23}$Na, $^{83}$Rb or $^{87}$Rb which have been experimentally studied, it is possible to realize an SLC in a lattice if the Hubbard interaction parameters $U_{2J}$ can be tuned properly through Feshbach resonances. Such an SLC can be distinguished from a spinor BEC or a Mott insulator experimentally. The superfluidity of SLC can be verified by single-atom-resolved imaging \cite{Sherson2010,Bakr2010}, or by seeing a Goldstone mode in the Bragg spectroscopy measurement \cite{Stamper-Kurn1999}. The spin rotational symmetry can be easily seen through a Stern-Gerlach imaging, since the symmetry ensures the populations on all spin $z$-components to be equal, namely, $\langle\hat{n}_{i,m}\rangle=\langle \hat{n}_i\rangle/5$ \cite{Lian2013}.

In summary, we have introduced the concept of SLC as a robust spin disordered charge condensate at zero temperature, and have showed how to realize it for spin $S\ge2$ bosons in a lattice. In particular we analyze in details the SLC phase for spin $2$ bosons, and verify that it has a spin gap and a gapless spinless Goldstone mode, which indicates a spin-charge separation. Bosons in the SLC move around in accompany with the creation and annihilation of spin singlet valence bonds, which is in analogy to electrons in the RVB state proposed for high temperature cuprates. Lastly, we shortly discussed the possibilities of SLC in the experiments.

\emph{Acknowledgements.} We acknowledge Hong Yao, Hui Zhai and Cenke Xu for helpful discussions. This work is supported by the NSF under grant numbers DMR-1305677.

\bibliography{SLC-ref}

\begin{thebibliography}{37}%
\makeatletter
\providecommand \@ifxundefined [1]{%
 \@ifx{#1\undefined}
}%
\providecommand \@ifnum [1]{%
 \ifnum #1\expandafter \@firstoftwo
 \else \expandafter \@secondoftwo
 \fi
}%
\providecommand \@ifx [1]{%
 \ifx #1\expandafter \@firstoftwo
 \else \expandafter \@secondoftwo
 \fi
}%
\providecommand \natexlab [1]{#1}%
\providecommand \enquote  [1]{``#1''}%
\providecommand \bibnamefont  [1]{#1}%
\providecommand \bibfnamefont [1]{#1}%
\providecommand \citenamefont [1]{#1}%
\providecommand \href@noop [0]{\@secondoftwo}%
\providecommand \href [0]{\begingroup \@sanitize@url \@href}%
\providecommand \@href[1]{\@@startlink{#1}\@@href}%
\providecommand \@@href[1]{\endgroup#1\@@endlink}%
\providecommand \@sanitize@url [0]{\catcode `\\12\catcode `\$12\catcode
  `\&12\catcode `\#12\catcode `\^12\catcode `\_12\catcode `\%12\relax}%
\providecommand \@@startlink[1]{}%
\providecommand \@@endlink[0]{}%
\providecommand \url  [0]{\begingroup\@sanitize@url \@url }%
\providecommand \@url [1]{\endgroup\@href {#1}{\urlprefix }}%
\providecommand \urlprefix  [0]{URL }%
\providecommand \Eprint [0]{\href }%
\providecommand \doibase [0]{http://dx.doi.org/}%
\providecommand \selectlanguage [0]{\@gobble}%
\providecommand \bibinfo  [0]{\@secondoftwo}%
\providecommand \bibfield  [0]{\@secondoftwo}%
\providecommand \translation [1]{[#1]}%
\providecommand \BibitemOpen [0]{}%
\providecommand \bibitemStop [0]{}%
\providecommand \bibitemNoStop [0]{.\EOS\space}%
\providecommand \EOS [0]{\spacefactor3000\relax}%
\providecommand \BibitemShut  [1]{\csname bibitem#1\endcsname}%
\let\auto@bib@innerbib\@empty
%</preamble>
\bibitem [{\citenamefont {Fisher}\ \emph {et~al.}(1989)\citenamefont {Fisher},
  \citenamefont {Weichman}, \citenamefont {Grinstein},\ and\ \citenamefont
  {Fisher}}]{Fisher1989}%
  \BibitemOpen
  \bibfield  {author} {\bibinfo {author} {\bibfnamefont {M.~P.~A.}\
  \bibnamefont {Fisher}}, \bibinfo {author} {\bibfnamefont {P.~B.}\
  \bibnamefont {Weichman}}, \bibinfo {author} {\bibfnamefont {G.}~\bibnamefont
  {Grinstein}}, \ and\ \bibinfo {author} {\bibfnamefont {D.~S.}\ \bibnamefont
  {Fisher}},\ } {\bibfield  {journal}
  {\bibinfo  {journal} {Phys. Rev. B}\ }\textbf {\bibinfo {volume} {40}},\
  \bibinfo {pages} {546} (\bibinfo {year} {1989})}\BibitemShut {NoStop}%
\bibitem [{\citenamefont {Stenger}\ \emph {et~al.}(1998)\citenamefont
  {Stenger}, \citenamefont {Inouye}, \citenamefont {Stamper-Kurn},
  \citenamefont {Miesner}, \citenamefont {Chikkatur},\ and\ \citenamefont
  {Ketterle}}]{Stenger1998}%
  \BibitemOpen
  \bibfield  {author} {\bibinfo {author} {\bibfnamefont {J.}~\bibnamefont
  {Stenger}}, \bibinfo {author} {\bibfnamefont {S.}~\bibnamefont {Inouye}},
  \bibinfo {author} {\bibfnamefont {D.~M.}\ \bibnamefont {Stamper-Kurn}},
  \bibinfo {author} {\bibfnamefont {H.-J.}\ \bibnamefont {Miesner}}, \bibinfo
  {author} {\bibfnamefont {A.~P.}\ \bibnamefont {Chikkatur}}, \ and\ \bibinfo
  {author} {\bibfnamefont {W.}~\bibnamefont {Ketterle}},\ } {\bibfield  {journal} {\bibinfo  {journal} {Nature}\ }\textbf
  {\bibinfo {volume} {396}},\ \bibinfo {pages} {345} (\bibinfo {year}
  {1998})}\BibitemShut {NoStop}%
\bibitem [{\citenamefont {Ho}(1998)}]{Ho1998}%
  \BibitemOpen
  \bibfield  {author} {\bibinfo {author} {\bibfnamefont {T.-L.}\ \bibnamefont
  {Ho}},\ } {\bibfield  {journal}
  {\bibinfo  {journal} {Phys. Rev. Lett.}\ }\textbf {\bibinfo {volume} {81}},\
  \bibinfo {pages} {742} (\bibinfo {year} {1998})}\BibitemShut {NoStop}%
\bibitem [{\citenamefont {Ohmi}\ and\ \citenamefont
  {Machida}(1998)}]{Ohmi1998}%
  \BibitemOpen
  \bibfield  {author} {\bibinfo {author} {\bibfnamefont {T.}~\bibnamefont
  {Ohmi}}\ and\ \bibinfo {author} {\bibfnamefont {K.}~\bibnamefont {Machida}},\
  } {\bibfield  {journal} {\bibinfo
  {journal} {J. Phys. Soc. Jpn.}\ }\textbf {\bibinfo {volume} {67}},\ \bibinfo
  {pages} {1822} (\bibinfo {year} {1998})}\BibitemShut {NoStop}%
\bibitem [{\citenamefont {Ciobanu}\ \emph {et~al.}(2000)\citenamefont
  {Ciobanu}, \citenamefont {Yip},\ and\ \citenamefont {Ho}}]{Ciobanu2000}%
  \BibitemOpen
  \bibfield  {author} {\bibinfo {author} {\bibfnamefont {C.~V.}\ \bibnamefont
  {Ciobanu}}, \bibinfo {author} {\bibfnamefont {S.-K.}\ \bibnamefont {Yip}}, \
  and\ \bibinfo {author} {\bibfnamefont {T.-L.}\ \bibnamefont {Ho}},\ } {\bibfield  {journal} {\bibinfo
  {journal} {Phys. Rev. A}\ }\textbf {\bibinfo {volume} {61}},\ \bibinfo
  {pages} {033607} (\bibinfo {year} {2000})}\BibitemShut {NoStop}%
\bibitem [{\citenamefont {Ueda}\ and\ \citenamefont {Koashi}(2002)}]{Ueda2002}%
  \BibitemOpen
  \bibfield  {author} {\bibinfo {author} {\bibfnamefont {M.}~\bibnamefont
  {Ueda}}\ and\ \bibinfo {author} {\bibfnamefont {M.}~\bibnamefont {Koashi}},\
  } {\bibfield  {journal} {\bibinfo
  {journal} {Phys. Rev. A}\ }\textbf {\bibinfo {volume} {65}},\ \bibinfo
  {pages} {063602} (\bibinfo {year} {2002})}\BibitemShut {NoStop}%
\bibitem [{\citenamefont {Barnett}\ \emph {et~al.}(2006)\citenamefont
  {Barnett}, \citenamefont {Turner},\ and\ \citenamefont
  {Demler}}]{Barnett2006}%
  \BibitemOpen
  \bibfield  {author} {\bibinfo {author} {\bibfnamefont {R.}~\bibnamefont
  {Barnett}}, \bibinfo {author} {\bibfnamefont {A.}~\bibnamefont {Turner}}, \
  and\ \bibinfo {author} {\bibfnamefont {E.}~\bibnamefont {Demler}},\ } {\bibfield  {journal} {\bibinfo
  {journal} {Phys. Rev. Lett.}\ }\textbf {\bibinfo {volume} {97}},\ \bibinfo
  {pages} {180412} (\bibinfo {year} {2006})}\BibitemShut {NoStop}%
\bibitem [{\citenamefont {Santos}\ and\ \citenamefont
  {Pfau}(2006)}]{Santos2006}%
  \BibitemOpen
  \bibfield  {author} {\bibinfo {author} {\bibfnamefont {L.}~\bibnamefont
  {Santos}}\ and\ \bibinfo {author} {\bibfnamefont {T.}~\bibnamefont {Pfau}},\
  } {\bibfield  {journal}
  {\bibinfo  {journal} {Phys. Rev. Lett.}\ }\textbf {\bibinfo {volume} {96}},\
  \bibinfo {pages} {190404} (\bibinfo {year} {2006})}\BibitemShut {NoStop}%
\bibitem [{\citenamefont {Diener}\ and\ \citenamefont {Ho}(2006)}]{Diener2006}%
  \BibitemOpen
  \bibfield  {author} {\bibinfo {author} {\bibfnamefont {R.~B.}\ \bibnamefont
  {Diener}}\ and\ \bibinfo {author} {\bibfnamefont {T.-L.}\ \bibnamefont
  {Ho}},\ } {\bibfield  {journal}
  {\bibinfo  {journal} {Phys. Rev. Lett.}\ }\textbf {\bibinfo {volume} {96}},\
  \bibinfo {pages} {190405} (\bibinfo {year} {2006})}\BibitemShut {NoStop}%
\bibitem [{\citenamefont {Kawaguchi}\ and\ \citenamefont
  {Ueda}(2011)}]{Kawaguchi2011}%
  \BibitemOpen
  \bibfield  {author} {\bibinfo {author} {\bibfnamefont {Y.}~\bibnamefont
  {Kawaguchi}}\ and\ \bibinfo {author} {\bibfnamefont {M.}~\bibnamefont
  {Ueda}},\ } {\bibfield  {journal}
  {\bibinfo  {journal} {Phys. Rev. A}\ }\textbf {\bibinfo {volume} {84}},\
  \bibinfo {pages} {053616} (\bibinfo {year} {2011})}\BibitemShut {NoStop}%
\bibitem [{\citenamefont {Lian}\ \emph {et~al.}(2012)\citenamefont {Lian},
  \citenamefont {Ho},\ and\ \citenamefont {Zhai}}]{Lian2012}%
  \BibitemOpen
  \bibfield  {author} {\bibinfo {author} {\bibfnamefont {B.}~\bibnamefont
  {Lian}}, \bibinfo {author} {\bibfnamefont {T.-L.}\ \bibnamefont {Ho}}, \ and\
  \bibinfo {author} {\bibfnamefont {H.}~\bibnamefont {Zhai}},\ } {\bibfield  {journal} {\bibinfo  {journal} {Phys.
  Rev. A}\ }\textbf {\bibinfo {volume} {85}},\ \bibinfo {pages} {051606}
  (\bibinfo {year} {2012})}\BibitemShut {NoStop}%
\bibitem [{\citenamefont {Mueller}\ \emph {et~al.}(2006)\citenamefont
  {Mueller}, \citenamefont {Ho}, \citenamefont {Ueda},\ and\ \citenamefont
  {Baym}}]{Mueller2006}%
  \BibitemOpen
  \bibfield  {author} {\bibinfo {author} {\bibfnamefont {E.~J.}\ \bibnamefont
  {Mueller}}, \bibinfo {author} {\bibfnamefont {T.-L.}\ \bibnamefont {Ho}},
  \bibinfo {author} {\bibfnamefont {M.}~\bibnamefont {Ueda}}, \ and\ \bibinfo
  {author} {\bibfnamefont {G.}~\bibnamefont {Baym}},\ } {\bibfield  {journal} {\bibinfo  {journal} {Phys.
  Rev. A}\ }\textbf {\bibinfo {volume} {74}},\ \bibinfo {pages} {033612}
  (\bibinfo {year} {2006})}\BibitemShut {NoStop}%
\bibitem [{\citenamefont {Law}\ \emph {et~al.}(1998)\citenamefont {Law},
  \citenamefont {Pu},\ and\ \citenamefont {Bigelow}}]{Law1998}%
  \BibitemOpen
  \bibfield  {author} {\bibinfo {author} {\bibfnamefont {C.~K.}\ \bibnamefont
  {Law}}, \bibinfo {author} {\bibfnamefont {H.}~\bibnamefont {Pu}}, \ and\
  \bibinfo {author} {\bibfnamefont {N.~P.}\ \bibnamefont {Bigelow}},\ } {\bibfield  {journal} {\bibinfo
  {journal} {Phys. Rev. Lett.}\ }\textbf {\bibinfo {volume} {81}},\ \bibinfo
  {pages} {5257} (\bibinfo {year} {1998})}\BibitemShut {NoStop}%
\bibitem [{\citenamefont {Ho}\ and\ \citenamefont {Yip}(2000)}]{Ho2000b}%
  \BibitemOpen
  \bibfield  {author} {\bibinfo {author} {\bibfnamefont {T.-L.}\ \bibnamefont
  {Ho}}\ and\ \bibinfo {author} {\bibfnamefont {S.~K.}\ \bibnamefont {Yip}},\
  } {\bibfield  {journal} {\bibinfo
   {journal} {Phys. Rev. Lett.}\ }\textbf {\bibinfo {volume} {84}},\ \bibinfo
  {pages} {4031} (\bibinfo {year} {2000})}\BibitemShut {NoStop}%
\bibitem [{\citenamefont {Mukerjee}\ \emph {et~al.}(2006)\citenamefont
  {Mukerjee}, \citenamefont {Xu},\ and\ \citenamefont {Moore}}]{Mukerjee2006}%
  \BibitemOpen
  \bibfield  {author} {\bibinfo {author} {\bibfnamefont {S.}~\bibnamefont
  {Mukerjee}}, \bibinfo {author} {\bibfnamefont {C.}~\bibnamefont {Xu}}, \ and\
  \bibinfo {author} {\bibfnamefont {J.~E.}\ \bibnamefont {Moore}},\ } {\bibfield  {journal} {\bibinfo
  {journal} {Phys. Rev. Lett.}\ }\textbf {\bibinfo {volume} {97}},\ \bibinfo
  {pages} {120406} (\bibinfo {year} {2006})}\BibitemShut {NoStop}%
\bibitem [{\citenamefont {Natu}\ and\ \citenamefont
  {Mueller}(2011)}]{Natu2011}%
  \BibitemOpen
  \bibfield  {author} {\bibinfo {author} {\bibfnamefont {S.~S.}\ \bibnamefont
  {Natu}}\ and\ \bibinfo {author} {\bibfnamefont {E.~J.}\ \bibnamefont
  {Mueller}},\ } {\bibfield
  {journal} {\bibinfo  {journal} {Phys. Rev. A}\ }\textbf {\bibinfo {volume}
  {84}},\ \bibinfo {pages} {053625} (\bibinfo {year} {2011})}\BibitemShut
  {NoStop}%
\bibitem [{\citenamefont {Demler}\ and\ \citenamefont
  {Zhou}(2002)}]{Demler2002}%
  \BibitemOpen
  \bibfield  {author} {\bibinfo {author} {\bibfnamefont {E.}~\bibnamefont
  {Demler}}\ and\ \bibinfo {author} {\bibfnamefont {F.}~\bibnamefont {Zhou}},\
  } {\bibfield  {journal}
  {\bibinfo  {journal} {Phys. Rev. Lett.}\ }\textbf {\bibinfo {volume} {88}},\
  \bibinfo {pages} {163001} (\bibinfo {year} {2002})}\BibitemShut {NoStop}%
\bibitem [{\citenamefont {Zhou}\ and\ \citenamefont {Snoek}(2003)}]{Zhou2003}%
  \BibitemOpen
  \bibfield  {author} {\bibinfo {author} {\bibfnamefont {F.}~\bibnamefont
  {Zhou}}\ and\ \bibinfo {author} {\bibfnamefont {M.}~\bibnamefont {Snoek}},\
  } {\bibfield
  {journal} {\bibinfo  {journal} {Ann. Phys.}\ }\textbf {\bibinfo {volume}
  {308}},\ \bibinfo {pages} {692 } (\bibinfo {year} {2003})}\BibitemShut
  {NoStop}%
\bibitem [{\citenamefont {Ruostekoski}\ and\ \citenamefont
  {Dutton}(2007)}]{Ruostekoski2007}%
  \BibitemOpen
  \bibfield  {author} {\bibinfo {author} {\bibfnamefont {J.}~\bibnamefont
  {Ruostekoski}}\ and\ \bibinfo {author} {\bibfnamefont {Z.}~\bibnamefont
  {Dutton}},\ } {\bibfield
  {journal} {\bibinfo  {journal} {Phys. Rev. A}\ }\textbf {\bibinfo {volume}
  {76}},\ \bibinfo {pages} {063607} (\bibinfo {year} {2007})}\BibitemShut
  {NoStop}%
\bibitem [{\citenamefont {Balents}(2010)}]{Balents2010}%
  \BibitemOpen
  \bibfield  {author} {\bibinfo {author} {\bibfnamefont {L.}~\bibnamefont
  {Balents}},\ } {\bibfield  {journal}
  {\bibinfo  {journal} {Nature}\ }\textbf {\bibinfo {volume} {464}},\ \bibinfo
  {pages} {199} (\bibinfo {year} {2010})}\BibitemShut {NoStop}%
\bibitem [{\citenamefont {Anderson}(1987)}]{Anderson1987}%
  \BibitemOpen
  \bibfield  {author} {\bibinfo {author} {\bibfnamefont {P.~W.}\ \bibnamefont
  {Anderson}},\ } {\bibfield
  {journal} {\bibinfo  {journal} {Science}\ }\textbf {\bibinfo {volume}
  {235}},\ \bibinfo {pages} {1196} (\bibinfo {year} {1987})}\BibitemShut
  {NoStop}%
\bibitem [{\citenamefont {Kivelson}\ \emph {et~al.}(1987)\citenamefont
  {Kivelson}, \citenamefont {Rokhsar},\ and\ \citenamefont
  {Sethna}}]{Kivelson1987}%
  \BibitemOpen
  \bibfield  {author} {\bibinfo {author} {\bibfnamefont {S.~A.}\ \bibnamefont
  {Kivelson}}, \bibinfo {author} {\bibfnamefont {D.~S.}\ \bibnamefont
  {Rokhsar}}, \ and\ \bibinfo {author} {\bibfnamefont {J.~P.}\ \bibnamefont
  {Sethna}},\ } {\bibfield  {journal}
  {\bibinfo  {journal} {Phys. Rev. B}\ }\textbf {\bibinfo {volume} {35}},\
  \bibinfo {pages} {8865} (\bibinfo {year} {1987})}\BibitemShut {NoStop}%
\bibitem [{sup()}]{supplement}%
  \BibitemOpen
  \href@noop {} {}\bibinfo {note} {See Supplemental Online Material for
  details.}\BibitemShut {Stop}%
\bibitem [{\citenamefont {Snoek}\ \emph {et~al.}(2009)\citenamefont {Snoek},
  \citenamefont {Song},\ and\ \citenamefont {Zhou}}]{Snoek2009}%
  \BibitemOpen
  \bibfield  {author} {\bibinfo {author} {\bibfnamefont {M.}~\bibnamefont
  {Snoek}}, \bibinfo {author} {\bibfnamefont {J.~L.}\ \bibnamefont {Song}}, \
  and\ \bibinfo {author} {\bibfnamefont {F.}~\bibnamefont {Zhou}},\ } {\bibfield  {journal} {\bibinfo
  {journal} {Phys. Rev. A}\ }\textbf {\bibinfo {volume} {80}},\ \bibinfo
  {pages} {053618} (\bibinfo {year} {2009})}\BibitemShut {NoStop}%
\bibitem [{\citenamefont {Zhou}\ and\ \citenamefont
  {Semenoff}(2006)}]{Zhou2006}%
  \BibitemOpen
  \bibfield  {author} {\bibinfo {author} {\bibfnamefont {F.}~\bibnamefont
  {Zhou}}\ and\ \bibinfo {author} {\bibfnamefont {G.~W.}\ \bibnamefont
  {Semenoff}},\ } {\bibfield
  {journal} {\bibinfo  {journal} {Phys. Rev. Lett.}\ }\textbf {\bibinfo
  {volume} {97}},\ \bibinfo {pages} {180411} (\bibinfo {year}
  {2006})}\BibitemShut {NoStop}%
\bibitem [{\citenamefont {Lian}\ and\ \citenamefont {Zhang}(2014)}]{Lian2013}%
  \BibitemOpen
  \bibfield  {author} {\bibinfo {author} {\bibfnamefont {B.}~\bibnamefont
  {Lian}}\ and\ \bibinfo {author} {\bibfnamefont {S.}~\bibnamefont {Zhang}},\
  } {\bibfield  {journal} {\bibinfo
  {journal} {Phys. Rev. B}\ }\textbf {\bibinfo {volume} {89}},\ \bibinfo
  {pages} {041110} (\bibinfo {year} {2014})}\BibitemShut {NoStop}%
\bibitem [{\citenamefont {Yang}(1962)}]{Yang1962}%
  \BibitemOpen
  \bibfield  {author} {\bibinfo {author} {\bibfnamefont {C.~N.}\ \bibnamefont
  {Yang}},\ } {\bibfield  {journal}
  {\bibinfo  {journal} {Rev. Mod. Phys.}\ }\textbf {\bibinfo {volume} {34}},\
  \bibinfo {pages} {694} (\bibinfo {year} {1962})}\BibitemShut {NoStop}%
\bibitem [{\citenamefont {Feynman}(1953)}]{Feynman1953}%
  \BibitemOpen
  \bibfield  {author} {\bibinfo {author} {\bibfnamefont {R.~P.}\ \bibnamefont
  {Feynman}},\ } {\bibfield  {journal}
  {\bibinfo  {journal} {Phys. Rev.}\ }\textbf {\bibinfo {volume} {91}},\
  \bibinfo {pages} {1291} (\bibinfo {year} {1953})}\BibitemShut {NoStop}%
\bibitem [{\citenamefont {Anderson}(1973)}]{Anderson1973}%
  \BibitemOpen
  \bibfield  {author} {\bibinfo {author} {\bibfnamefont {P.~W.}\ \bibnamefont
  {Anderson}},\ } {\bibfield  {journal}
  {\bibinfo  {journal} {Mat. Res. Bull.}\ }\textbf {\bibinfo {volume} {8}},\
  \bibinfo {pages} {153 } (\bibinfo {year} {1973})}\BibitemShut {NoStop}%
\bibitem [{\citenamefont {Sutherland}(1988)}]{Sutherland1988}%
  \BibitemOpen
  \bibfield  {author} {\bibinfo {author} {\bibfnamefont {B.}~\bibnamefont
  {Sutherland}},\ } {\bibfield
  {journal} {\bibinfo  {journal} {Phys. Rev. B}\ }\textbf {\bibinfo {volume}
  {37}},\ \bibinfo {pages} {3786} (\bibinfo {year} {1988})}\BibitemShut
  {NoStop}%
\bibitem [{\citenamefont {Greiner}\ \emph {et~al.}(2002)\citenamefont
  {Greiner}, \citenamefont {Mandel}, \citenamefont {Esslinger}, \citenamefont
  {Hansch},\ and\ \citenamefont {Bloch}}]{Greiner2002}%
  \BibitemOpen
  \bibfield  {author} {\bibinfo {author} {\bibfnamefont {M.}~\bibnamefont
  {Greiner}}, \bibinfo {author} {\bibfnamefont {O.}~\bibnamefont {Mandel}},
  \bibinfo {author} {\bibfnamefont {T.}~\bibnamefont {Esslinger}}, \bibinfo
  {author} {\bibfnamefont {T.~W.}\ \bibnamefont {Hansch}}, \ and\ \bibinfo
  {author} {\bibfnamefont {I.}~\bibnamefont {Bloch}},\ } {\bibfield  {journal} {\bibinfo  {journal} {Nature}\
  }\textbf {\bibinfo {volume} {415}},\ \bibinfo {pages} {39} (\bibinfo {year}
  {2002})}\BibitemShut {NoStop}%
\bibitem [{\citenamefont {Gerbier}\ \emph {et~al.}(2005)\citenamefont
  {Gerbier}, \citenamefont {Widera}, \citenamefont {F\"olling}, \citenamefont
  {Mandel}, \citenamefont {Gericke},\ and\ \citenamefont
  {Bloch}}]{Gerbier2005}%
  \BibitemOpen
  \bibfield  {author} {\bibinfo {author} {\bibfnamefont {F.}~\bibnamefont
  {Gerbier}}, \bibinfo {author} {\bibfnamefont {A.}~\bibnamefont {Widera}},
  \bibinfo {author} {\bibfnamefont {S.}~\bibnamefont {F\"olling}}, \bibinfo
  {author} {\bibfnamefont {O.}~\bibnamefont {Mandel}}, \bibinfo {author}
  {\bibfnamefont {T.}~\bibnamefont {Gericke}}, \ and\ \bibinfo {author}
  {\bibfnamefont {I.}~\bibnamefont {Bloch}},\ } {\bibfield  {journal} {\bibinfo  {journal}
  {Phys. Rev. Lett.}\ }\textbf {\bibinfo {volume} {95}},\ \bibinfo {pages}
  {050404} (\bibinfo {year} {2005})}\BibitemShut {NoStop}%
\bibitem [{\citenamefont {St\"oferle}\ \emph {et~al.}(2004)\citenamefont
  {St\"oferle}, \citenamefont {Moritz}, \citenamefont {Schori}, \citenamefont
  {K\"ohl},\ and\ \citenamefont {Esslinger}}]{Stoferle2004}%
  \BibitemOpen
  \bibfield  {author} {\bibinfo {author} {\bibfnamefont {T.}~\bibnamefont
  {St\"oferle}}, \bibinfo {author} {\bibfnamefont {H.}~\bibnamefont {Moritz}},
  \bibinfo {author} {\bibfnamefont {C.}~\bibnamefont {Schori}}, \bibinfo
  {author} {\bibfnamefont {M.}~\bibnamefont {K\"ohl}}, \ and\ \bibinfo {author}
  {\bibfnamefont {T.}~\bibnamefont {Esslinger}},\ } {\bibfield  {journal} {\bibinfo  {journal}
  {Phys. Rev. Lett.}\ }\textbf {\bibinfo {volume} {92}},\ \bibinfo {pages}
  {130403} (\bibinfo {year} {2004})}\BibitemShut {NoStop}%
\bibitem [{\citenamefont {Spielman}\ \emph {et~al.}(2007)\citenamefont
  {Spielman}, \citenamefont {Phillips},\ and\ \citenamefont
  {Porto}}]{Spielman2007}%
  \BibitemOpen
  \bibfield  {author} {\bibinfo {author} {\bibfnamefont {I.~B.}\ \bibnamefont
  {Spielman}}, \bibinfo {author} {\bibfnamefont {W.~D.}\ \bibnamefont
  {Phillips}}, \ and\ \bibinfo {author} {\bibfnamefont {J.~V.}\ \bibnamefont
  {Porto}},\ } {\bibfield
  {journal} {\bibinfo  {journal} {Phys. Rev. Lett.}\ }\textbf {\bibinfo
  {volume} {98}},\ \bibinfo {pages} {080404} (\bibinfo {year}
  {2007})}\BibitemShut {NoStop}%
\bibitem [{\citenamefont {Sherson}\ \emph {et~al.}(2010)\citenamefont
  {Sherson}, \citenamefont {Weitenberg}, \citenamefont {Endres}, \citenamefont
  {Cheneau}, \citenamefont {Bloch},\ and\ \citenamefont {Kuhr}}]{Sherson2010}%
  \BibitemOpen
  \bibfield  {author} {\bibinfo {author} {\bibfnamefont {J.~F.}\ \bibnamefont
  {Sherson}}, \bibinfo {author} {\bibfnamefont {C.}~\bibnamefont {Weitenberg}},
  \bibinfo {author} {\bibfnamefont {M.}~\bibnamefont {Endres}}, \bibinfo
  {author} {\bibfnamefont {M.}~\bibnamefont {Cheneau}}, \bibinfo {author}
  {\bibfnamefont {I.}~\bibnamefont {Bloch}}, \ and\ \bibinfo {author}
  {\bibfnamefont {S.}~\bibnamefont {Kuhr}},\ } {\bibfield  {journal} {\bibinfo  {journal} {Nature}\
  }\textbf {\bibinfo {volume} {467}},\ \bibinfo {pages} {68} (\bibinfo {year}
  {2010})}\BibitemShut {NoStop}%
\bibitem [{\citenamefont {Bakr}\ \emph {et~al.}(2010)\citenamefont {Bakr},
  \citenamefont {Peng}, \citenamefont {Tai}, \citenamefont {Ma}, \citenamefont
  {Simon}, \citenamefont {Gillen}, \citenamefont {F\"olling}, \citenamefont
  {Pollet},\ and\ \citenamefont {Greiner}}]{Bakr2010}%
  \BibitemOpen
  \bibfield  {author} {\bibinfo {author} {\bibfnamefont {W.~S.}\ \bibnamefont
  {Bakr}}, \bibinfo {author} {\bibfnamefont {A.}~\bibnamefont {Peng}}, \bibinfo
  {author} {\bibfnamefont {M.~E.}\ \bibnamefont {Tai}}, \bibinfo {author}
  {\bibfnamefont {R.}~\bibnamefont {Ma}}, \bibinfo {author} {\bibfnamefont
  {J.}~\bibnamefont {Simon}}, \bibinfo {author} {\bibfnamefont {J.~I.}\
  \bibnamefont {Gillen}}, \bibinfo {author} {\bibfnamefont {S.}~\bibnamefont
  {F\"olling}}, \bibinfo {author} {\bibfnamefont {L.}~\bibnamefont {Pollet}}, \
  and\ \bibinfo {author} {\bibfnamefont {M.}~\bibnamefont {Greiner}},\ } {\bibfield  {journal} {\bibinfo  {journal}
  {Science}\ }\textbf {\bibinfo {volume} {329}},\ \bibinfo {pages} {547}
  (\bibinfo {year} {2010})}\BibitemShut {NoStop}%
\bibitem [{\citenamefont {Stamper-Kurn}\ \emph {et~al.}(1999)\citenamefont
  {Stamper-Kurn}, \citenamefont {Chikkatur}, \citenamefont {G\"orlitz},
  \citenamefont {Inouye}, \citenamefont {Gupta}, \citenamefont {Pritchard},\
  and\ \citenamefont {Ketterle}}]{Stamper-Kurn1999}%
  \BibitemOpen
  \bibfield  {author} {\bibinfo {author} {\bibfnamefont {D.~M.}\ \bibnamefont
  {Stamper-Kurn}}, \bibinfo {author} {\bibfnamefont {A.~P.}\ \bibnamefont
  {Chikkatur}}, \bibinfo {author} {\bibfnamefont {A.}~\bibnamefont
  {G\"orlitz}}, \bibinfo {author} {\bibfnamefont {S.}~\bibnamefont {Inouye}},
  \bibinfo {author} {\bibfnamefont {S.}~\bibnamefont {Gupta}}, \bibinfo
  {author} {\bibfnamefont {D.~E.}\ \bibnamefont {Pritchard}}, \ and\ \bibinfo
  {author} {\bibfnamefont {W.}~\bibnamefont {Ketterle}},\ } {\bibfield  {journal} {\bibinfo  {journal}
  {Phys. Rev. Lett.}\ }\textbf {\bibinfo {volume} {83}},\ \bibinfo {pages}
  {2876} (\bibinfo {year} {1999})}\BibitemShut {NoStop}%
\end{thebibliography}%

\begin{widetext}

\section*{Supplementary material}

\subsection{On-site state energy spectrums of spin $1$ bosons}
The on-site interaction energy of spin $1$ bosons can be rewritten as
\begin{equation}
H_I=\sum_i\left[-\mu\hat{n}_i+f_0\hat{n}_i(\hat{n}_i-1)+f_1\hat{\mathbf{S}}_i^2\right]\ ,
\end{equation}
where we have redefined interaction parameters $f_0=(2U_2+U_0)/6$ and $f_1=(U_2-U_0)/6$, while $\hat{\mathbf{S}}_i=\psi^\dag_{i,a}\mathbf{S}_{ab}\psi_{i,b}$ is the total spin operator on site $i$ ($\mathbf{S}_{ab}$ is the spin $1$ matrix). The on-site states of spin $1$ bosons can then be labeled uniquely as $|n,l,m\rangle$, where $n$ is the number of bosons, $l$ is the total angular momentum, while $m$ is the $z$-component of the total angular momentum. The values of $n$ and $l$ are restricted to satisfy $n\ge l\ge0$ and $n+l$ even. The energy spectrum of the spin $1$ boson on-site state $|n,l,m\rangle$ is then
\begin{equation}
E_{nl}(S=1)=-\mu n+f_0n(n-1)+f_1l(l+1)\ .
\end{equation}
If one want to make both the lowest two energy levels be spin singlets with $l=0$, one has to first tune the chemical potential to $\mu=(4p-3)f_0$ where $p$ is a positive integer, so that we have two low energy singlet states $|2p-2,0,0\rangle$ and $|2p,0,0\rangle$ degenerate. Then one has to raise the energy of states $|2p-1,1,m\rangle$ to make the two singlet states the lowest. This means
\begin{equation}
E_{2p-1,1}(S=1)-E_{2p,0}(S=1)=2f_1-f_0=-U_0/2>0\ ,
\end{equation}
namely $U_0<0$. Such an attractive interaction between bosons is unlikely to be achieved experimentally, and may induce instabilities against decaying mechanisms like the pair formation. This is the reason we require all $U_{2J}\ge0$ at the beginning. Therefore, it is impossible to find a physical parameter regime where the SLC could arise for spin $1$ bosons. One can still study such spin $1$ models with attractive interactions from a purely theoretical perspective, which may yield an SLC with elementary charge $2$, yet we will not discuss this case here. Furthermore, such a charge $2$ SLC may be in some sense viewed as a traditional BEC of the binary molecules formed by two bosons under the attractive $U_0$, and is therefore not as interesting as the charge $1$ SLC of spin $2$ bosons here.

\subsection{On-site state energy spectrums of spin $2$ bosons}

Now we briefly introduce the more complicated on-site state spectrum of spin $2$ bosons. This is discussed in Ref. \cite{Zhou2006,Ueda2002}, and in more details in Ref. \cite{Snoek2009}. In general, unlike those of spin $1$ bosons, the on-site states of bosons of higher spin cannot be uniquely labeled by $n$, $l$ and $m$. One has to introduce additional quantum numbers to label the states.

For spin $2$ bosons, we can define two spinless operators $\hat{D}^\dag_i=(1/\sqrt{40})tr(\psi_i^\dag\psi_i^\dag)$ and $\hat{T}^\dag_i=(1/\sqrt{140})tr(\psi_i^\dag\psi_i^\dag\psi_i^\dag)$, where the $\psi_i^\dag$ is the rewritten traceless matrix ($2$-tensor) form of field operators $\psi_{i,m}^\dag$ as is defined in Eq. (1) of Ref. \cite{Zhou2006}. Their explicit forms are given by
\begin{equation}
\begin{split}
&\hat{D}^\dag_i=\frac{1}{\sqrt{10}}\left(2\psi^\dag_{i,+2}\psi^\dag_{i,-2}-2\psi^\dag_{i,+1}\psi^\dag_{i,-1}+{\psi^{\dag}_{i,0}}^2\right)\ ,\\ &\hat{T}^\dag_i=\frac{1}{\sqrt{420}} \left(12\psi^\dag_{i,+2}\psi^\dag_{i,0}\psi^\dag_{i,-2}+6\psi^{\dag}_{i,+1}\psi^\dag_{i,0}\psi^{\dag}_{i,-1}-3\sqrt{6}\psi^\dag_{i,+2} {\psi^\dag_{i,-1}}^2 -3\sqrt{6}\psi^\dag_{i,-2}{\psi^\dag_{i,+1}}^2-2{\psi_{i,0}^\dag}^3\right)\ .
\end{split}
\end{equation}
They satisfy the relations $[\hat{D}^\dag_i\hat{D}_i,H_I]=[\hat{T}^\dag_i\hat{T}_i,H_I]=0$. The interaction Hamiltonian can be rewritten in the form
\begin{equation}
H_I=\sum_i\left[-\mu\hat{n}_i+\frac{g_0}{2}\hat{n}_i(\hat{n}_i-1)+\frac{g_1}{2}(\hat{\mathbf{S}}_i^2-6\hat{n}_i)+5g_2\hat{D}^\dag_i\hat{D}_i\right]\ ,
\end{equation}
where $g_0=(4U_2+3U_4)/7$, $g_1=(U_4-U_2)/7$, $g_2=(U_0-U_4)/5-2(U_2-U_4)/7$, while $\hat{\mathbf{S}}_i=\psi^\dag_{i,a}\mathbf{S}_{ab}\psi_{i,b}$ is the total spin operator for spin $2$ bosons, with $\mathbf{S}_{ab}$ here the spin $2$ matrix \cite{Snoek2009,Zhou2006}. The key to find the energy spectrum is to define three operators $\hat{D}^+_i=\sqrt{5/2}\hat{D}^\dag_i$, $\hat{D}^-_i=\sqrt{5/2}\hat{D}_i$ and $\hat{D}^z_i=(2\hat{n}_i+5)/4$, and note that they satisfy the $SU(1,1)$ algebra:
\begin{equation}
[\hat{D}^+_i,\hat{D}^-_i]=-2\hat{D}^z_i\ ,\qquad\qquad [\hat{D}^z_i,\hat{D}^\pm_i]=\pm\hat{D}^\pm_i\ .
\end{equation}
They give us a Casimir operator $\hat{\mathbf{D}}_i^2=-(\hat{D}^-_i\hat{D}^+_i+\hat{D}^+_i\hat{D}^-_i)/2+\hat{D}^{z2}_i$. Any on-site state can therefore be written in the form $|\chi\rangle_i=(\hat{D}_i^\dag)^k|\chi_0\rangle_i$, where $|\chi_0\rangle_i$ is a state satisfying $\hat{D}_i|\chi_0\rangle_i=0$. Suppose the state $|\chi_0\rangle_i$ satisfies $\hat{n}_i|\chi_0\rangle_i=\gamma|\chi_0\rangle_i$ where $\gamma$ is an integer. Then a general on-site state can be labeled by $|\chi\rangle_i=|n,l,m,\gamma\rangle_i=(\hat{D}_i^\dag)^{(n-\gamma)/2}|\chi_0\rangle_i$, and its energy is given by \cite{Snoek2009}
\begin{equation}
E_{nl\gamma}(S=2)=-\mu n+\frac{g_0}{2}n(n-1)+\frac{g_1}{2}\left[l(l+1)-6n\right]+\frac{g_2}{8}\left[(2n+3)^2-(2\gamma+3)^2\right]\ .
\end{equation}
Certainly, there are some constraints between the values $n$, $l$ and $\gamma$, which is discussed in Ref. \cite{Snoek2009}.

For later use, we list in Tab. \ref{I} several of the on-site states (normalized) and their energies. For our purpose, it is sufficient to focus on only the lowest several on-site states shown in Fig. 1, which are distinguishable by $n$, $l$, $m$ solely, so we shall omit the quantum number $\gamma$ in most places for simplicity. In particular, $E_{42}$ in Fig. 1 denotes the energy of the state $|4,2,m,2\rangle$, since the state $|4,2,m,4\rangle$ does not take part in the three-time hopping process shown in Fig. 3(a) (there is no direct hopping from $|4,2,m,4\rangle$ to $|3,2,m',1\rangle$).

\begin{table}[t]
\caption{\label{I} Several normalized on-site states of spin $2$ bosons and their energies.}
\begin{ruledtabular}
\begin{tabular}{c|c|c}
quantum numbers $|n,l,m,\gamma\rangle$ & state construction (particle vacuum $|\Omega\rangle$) & on-site energy $E_{nl\gamma}$ \\
\colrule
$|1,2,m,1\rangle$ & $\psi^\dag_{i,m}|\Omega\rangle$ & $-\mu$ \\
$|2,0,0,0\rangle$ & $\hat{D}^\dag_i|\Omega\rangle$ & $U_0-2\mu$ \\
$|2,2,m,2\rangle$ & $(-1)^m\sqrt{\frac{5}{3}}\psi_{i,-m}\hat{T}^\dag_i|\Omega\rangle$ & $U_2-2\mu$ \\
$|2,4,m,2\rangle$ & $\frac{1}{\sqrt{2}}\langle4,m|2,m_1;2,m_2\rangle \psi_{m_1}^\dag\psi_{m_2}^\dag|\Omega\rangle$ & $U_4-2\mu$ \\
$|3,0,0,3\rangle$ & $\hat{T}^\dag_i|\Omega\rangle$ & $3U_2-3\mu$ \\
$|3,2,m,1\rangle$ & $\sqrt{\frac{5}{7}}\psi^\dag_{i,m}D^\dag|\Omega\rangle$ & $\frac{7}{5}U_0+\frac{4}{7}U_2+\frac{36}{35}U_4-3\mu$ \\
$|4,2,m,2\rangle$ & $\frac{\sqrt{5}}{3}\hat{D}^\dag_i|2,2,m,2\rangle$ & $\frac{9}{5}U_0+\frac{15}{7}U_2+\frac{72}{35}U_4-4\mu$ \\
$|4,2,m,4\rangle$ & $\sqrt{\frac{15}{22}}\psi^\dag_{i,m}\hat{T}_i^\dag|\Omega\rangle-\frac{1}{\sqrt{11}}|4,2,m,2\rangle$ & $\frac{33}{7}U_2+\frac{9}{7}U_4-4\mu$ \\
\end{tabular}
\end{ruledtabular}
\end{table}

By examining the on-site state energy spectrum of spin $2$ bosons, one finds when $0\le U_0<U_2<(36U_4+49U_0)/85$ and $\mu=\mu_0=3U_2-U_0$, the two spin singlet states $|2,0,0,0\rangle$ (dimer state) and $|3,0,0,3\rangle$ (trimer state) become degenerate and have the lowest energy.

\subsection{On the order parameter $\lambda$}
We have defined a spinless order parameter $\lambda$ in the main text:
\begin{equation}
\lambda=\langle\hat{\lambda}_i\rangle=\langle\sum_{m,m'=-2}^{2}D_{m'm}\psi_{i,m'}\psi_{i,m}\psi^\dag_{i,m+m'}\rangle\ ,
\end{equation}
where the coefficient $D_{m'm}=(-1)^{m+m'}\langle 0,0|2,-m-m';2,m+m'\rangle\langle2,m+m'|2,m';2,m\rangle$. To see $\lambda$ has a spin rotational symmetry, we first note that the three combinations:
$$\sum_{m}|2,m\rangle\psi_{i,m},\qquad \sum_{m}(-1)^{m}|2,-m\rangle\psi^\dag_{i,m}, \qquad\sum_{m}\left|2,m\rangle\langle 2,m\right|\ ,$$
are spin rotationally invariant as is easily verified, where $|2,m\rangle$ stands for a spin state with spin $2$. We can therefore rewrite the order parameter as
\begin{equation}
\lambda=\langle\Big(\langle 0,0|\Big) \left[\left(\sum_{m_1}\left|2,m_1\rangle\langle 2,m_1\right|\right) \left(\sum_{m_2}|2,m_2\rangle\psi_{i,m_2}\right)\left(\sum_{m_3}|2,m_3\rangle\psi_{i,m_3}\right)\right] \left(\sum_{m_4}(-1)^{m_4}|2,-m_4\rangle\psi^\dag_{i,m_4}\right)\rangle\ ,
\end{equation}
which is explicitly spin rotationally invariant. One can further verify directly that $\hat{\lambda}_i$ satisfies $[\hat{\mathbf{S}}_i,\hat{\lambda}_i]=0$ where $\hat{\mathbf{S}}_i$ is the total spin defined in last section, and $[\hat{n}_i,\hat{\lambda}_i]=-\hat{\lambda}_i$. Under U($1$) transformations $\psi_{i,m}\rightarrow e^{i\phi}\psi_{i,m}$, $\psi_{i,m}^\dag\rightarrow e^{-i\phi}\psi^\dag_{i,m}$, it is easy to see $\hat{\lambda}_i\rightarrow e^{i\phi}\hat{\lambda}_i$, so the order parameter carries charge $1$. One can therefore create a vortex around which the phase $\phi$ increases $2\pi M$ (required by periodic boundary condition), and if one measure the order parameter $\lambda$, there will be a $2\pi M$ flux in the vortex ($M\in\mathbb{Z}$).

We can define the 3-particle density matrix $\rho^{(3)}$ as
\begin{equation}
\rho^{(3)}_{m_1m_2m_3m_4m_5m_6}(\mathbf{x}_{i_1},\mathbf{x}_{i_2},\mathbf{x}_{i_3},\mathbf{x}_{i_4},\mathbf{x}_{i_5},\mathbf{x}_{i_6}) =\langle \psi^\dag_{i_1,m_1}\psi^\dag_{i_2,m_2}\psi^\dag_{i_3,m_3}\psi_{i_4,m_4}\psi_{i_5,m_5}\psi_{i_6,m_6}\rangle\ .
\end{equation}
The off diagonal terms of $\rho^{(3)}$ can be seen to decay exponentially as the six points $\mathbf{x}_{i_\alpha}$ are far from each other, unless three points of them coincide while the other three points also coincide. For instance, when $\mathbf{x}_{i_1}=\mathbf{x}_{i_2}=\mathbf{x}_{i_4}$ and $\mathbf{x}_{i_3}=\mathbf{x}_{i_5}=\mathbf{x}_{i_6}$, the matrix element is a constant $\sim |\lambda|^2$ as $|\mathbf{x}_{i_1}-\mathbf{x}_{i_3}|\rightarrow \infty$. Actually, a direct calculation using the mean-field variational wave function $|SLC\rangle$ gives
\begin{equation}
\begin{split}
&\rho^{(3)}_{m_1m_2m_3m_4m_5m_6}(\mathbf{x}_{i_1},\mathbf{x}_{i_2},\mathbf{x}_{i_3},\mathbf{x}_{i_4},\mathbf{x}_{i_5},\mathbf{x}_{i_6})\approx
\Big\{\delta_{m_1-m_4}\delta_{m_2-m_5}\delta_{m_3-m_6}\langle \hat{n}_i\rangle^3 e^{-(|\mathbf{x}_{i_1}-\mathbf{x}_{i_4}|+|\mathbf{x}_{i_2}-\mathbf{x}_{i_5}|+|\mathbf{x}_{i_3}-\mathbf{x}_{i_6}|)/\xi_S}\\
&+\sum_\mathcal{P\neq\mathbf{1}}\Big[(1,2,3)\leftrightarrow\mathcal{P}(1,2,3)\Big]\Big\}+ \Big\{ e^{-\left(|\mathbf{x}^{(1)}_{F}-\mathbf{x}_{i_1}|+|\mathbf{x}^{(1)}_{F}-\mathbf{x}_{i_2}|+|\mathbf{x}^{(1)}_{F}-\mathbf{x}_{i_4}|+ |\mathbf{x}^{(2)}_{F}-\mathbf{x}_{i_3}|+|\mathbf{x}^{(2)}_{F}-\mathbf{x}_{i_5}|+|\mathbf{x}^{(2)}_{F}-\mathbf{x}_{i_6}|\right)/\xi_S} \\ &\times \delta_{m_1+m_2-m_4}\delta_{m_5+m_6-m_3}D_{m_1m_2}^*D_{m_5m_6}|\lambda|^2 +(1\leftrightarrow3)+(2\leftrightarrow3)+(4\leftrightarrow5)+ (4\leftrightarrow6)+(1\leftrightarrow3,4\leftrightarrow5)\\
&+(1\leftrightarrow3,4\leftrightarrow6)+(2\leftrightarrow3,4\leftrightarrow5)+(2\leftrightarrow3,4\leftrightarrow6)\Big\}\ ,
\end{split}
\end{equation}
where $\mathcal{P}$ stands for permutation, while $\mathbf{x}^{(1)}_{F}$ and $\mathbf{x}^{(2)}_{F}$ are the Fermat points of triangles $\triangle\mathbf{x}_{i_1}\mathbf{x}_{i_2}\mathbf{x}_{i_4}$ and $\triangle\mathbf{x}_{i_3}\mathbf{x}_{i_5}\mathbf{x}_{i_6}$ respectively. The Fermat point of a triangle has a minimal sum of distances to the three vertices of the triangle. The first term represents the diagonal elements, while the second gives the off-diagonal elements. In particular, one sees that in the limit $|\mathbf{x}_{i}-\mathbf{x}_{j}|\rightarrow\infty$,
\begin{equation}
\rho^{(3)}_{m_1m_2m_3m_4m_5m_6}(\mathbf{x}_{i},\mathbf{x}_{i},\mathbf{x}_{j},\mathbf{x}_{i},\mathbf{x}_{j},\mathbf{x}_{j})
=\delta_{m_1+m_2-m_4}\delta_{m_5+m_6-m_3}D_{m_1m_2}^*D_{m_5m_6}|\lambda|^2\neq0\ .
\end{equation}
There is therefore an ODLRO in $\rho^{(3)}$, and the largest eigenvalue $r_3$ in $\rho^{(3)}$ is of order $N$. As is shown in the following section of supplementary material, there is therefore no ODLRO in $\rho^{(1)}$ and $\rho^{(2)}$.

This means that the bosons do not condense directly in the conventional way. As is seen from the order parameter $\lambda\sim\langle \psi\psi\psi^\dag\rangle$, any boson participating in the condensate is ``dressed": Its spin is fully screened by a local virtual particle-hole pair, while its charge remains unchanged since a particle-hole pair carries no charge. SLC is therefore a condensate of spinless ``dressed bosons".

\subsection{ODLRO in $k$-particle density matrix ($j>1$)}
In the definition of SLC, we require ODLRO in a $j$-particle density matrix $\rho^{(j)}$ ($j>1$), by which we mean there is no ODLRO in $k'$-particle density matrix $\rho^{(j')}$ if $j'<j$. We note that, the minimal condition for ODLRO to arise in the $j$-particle density matrix $\rho^{(j)}$ is to have the largest eigenvalue $r_j$ of $\rho^{(j)}$ of order $\mathcal{O}(N)$, where $N$ is the total number of particles \cite{Yang1962}. By Ref. \cite{Yang1962}, the largest eigenvalues $r_j$ of the reduced density matrices of bosons satisfy
\begin{equation}
r_1\le N\ ,\qquad r_1^2-r_1\le r_2\le N(N-1)\ ,\qquad r_1^3-2r_1^2-r_2\le r_3\le N(N-1)(N-2)\ ,\qquad \cdots
\end{equation}
So it is possible to have no eigenvalues of order $\mathcal{O}(N)$ in $\rho^{(j')}$ where $j'<j$, while having $r_j$ of order $\mathcal{O}(N)$. For fermions, the lowest density matrix for ODLRO to arise is the $2$-particle density matrix. For traditional BECs on a single-particle state, the largest eigenvalue $r^{(j)}$ of $\rho^{(j)}$ is of order $\mathcal{O}(N^j)$. In our SLC example of spin $2$ bosons here, we have a spinless order parameter $\lambda$ that is given by a three-boson operator, as is defined in Eq. (2) of the letter. This means the state has an ODLRO in the $3$-particle density matrix $\rho^{(3)}$, whose largest eigenvalue is given by
\begin{equation}
r_3\approx\sum_{i'}\langle\hat{\lambda}^\dag_i\hat{\lambda}_{i'}\rangle\approx\sum_{i'}\lambda^*(\mathbf{x}_i)\lambda(\mathbf{x}_{i'}) =N_S|\lambda|^2\sim \mathcal{O}(N)\ ,
\end{equation}
where $N_S$ is the number of sites. A straightforward corollary of this result is that the largest eigenvalue $r_1$ of the single-particle density matrix is no larger than $\mathcal{O}(N^{1/3})$, demonstrating that the SLC state is a non-SPS charge condensate. Similarly, the largest eigenvalue $r_2$ of $\rho^{(2)}$ is no larger than $\mathcal{O}(N^{2/3})$.

In contrast, the spin-paired condensate of spin $1$ bosons proposed in Ref. \cite{Law1998} has eigenvalues of order $N$ already in the single-particle density matrix $\rho^{(1)}$ \cite{Mueller2006}.

\subsection{Derivation and minimization of the variational energy $\mathcal{E}_G$}
The motivation of writing the trial wave function
\begin{equation}
|SLC\rangle=Sym\prod_{\langle ij\rangle}\Big[u+\sum_m\left(v\psi_{i,m}^\dag\psi_{j,m}+h.c.\right)\Big] \times\prod_i\Big(\alpha|2,0,0\rangle_i+\beta|3,0,0\rangle_i\Big)
\end{equation}
is very simple. Firstly, since we are in the Mott regime $t\ll U_{2J}$, the ground state should be a superposition of the dimer and trimer state, which are the only low energy on-site states. $H_t$ at this time only serves as a perturbation. As we have said, the superfluid necessarily needs the contribution of hopping $3$ times, we have to therefore write a more exact wave function corrected by the perturbation theory to include higher order perturbations, which can be approximately written in the above form.

We first derive the normalization condition of this mean-field variational wave function in the limit $v/u\ll1$, in which we have already assumed $|\alpha|^2+|\beta|^2=1$. This can be done more exactly via the loop gas approach discussed later, but a simple estimation up to the quadratic order $|v/u|^2$ is enough here. We first focus on a site $i$ and its $z$ neighbouring sites $j$. To the lowest order the wave function around site $i$ can be approximated as
\begin{equation}
|\phi(u,v)\rangle_i\approx u|s\rangle_i\prod_{j\in\langle ij\rangle}|s\rangle_j+v\psi^\dag_{i,m}|s\rangle_i\prod_{j\in\langle ij\rangle}\psi_{j,m}|s\rangle_j+v^*\psi_{i,m}|s\rangle_i\prod_{j\in\langle ij\rangle}\psi^\dag_{i,m}|s\rangle_j\ ,
\end{equation}
where we have written $|s\rangle_i=(\alpha|2,0,0\rangle_i+\beta|3,0,0\rangle_i)$ in short. The normalization of this wave function gives
\begin{equation}\label{nor}
1=\langle\phi(u,v)|\phi(u,v)\rangle_i\approx|u|^2+2z|v|^2\left({_i\langle} s|\psi_{i,m}\psi^\dag_{i,m}|s\rangle_i\right) \left({_j\langle} s|\psi^\dag_{j,m}\psi_{j,m}|s\rangle_j\right)=|u|^2+\frac{2}{5}z|v|^2(2+\beta^2)(7+\beta^2)\ .
\end{equation}
The entire wave function can be very roughly viewed as $N_S$ copies of state $|\phi(u,v)\rangle_i$, where $N_S$ is the number of sites. We can therefore coarsely estimate the normalization $\langle SLC|SLC\rangle\approx \langle\phi(u,v)|\phi(u,v)\rangle_i^{N_S/2}=1$, where the factor $1/2$ in the exponent is to counter the double counting of lattice bonds in estimating the hopping contributions in this approach. We shall therefore use Eq. (\ref{nor}) as the normalization constraint (whose exact form will not affect the phase diagram in the limit $v/u\rightarrow 0$).

Then we proceed to derive the variational energy per site $\mathcal{E}_G=\langle H_I+H_t\rangle/N_S$. For convenience, we shift the zero point of the on-site interaction energy in the following so that the energy of the dimer state $|2,0,0\rangle$ is fixed at $0$. Assume the chemical potential is $\mu=\mu_0-\Delta$ where $\Delta$ is small. It is easy to see that $E_{30}-E_{20}=\Delta$. Following the above, we want to keep only up to the quadratic order $|v/u|^2$.

The interaction energy comes from two parts: the on-site energy of the singlet state $|s\rangle_i$, and the energy of singlet valence bonds arising from the background. As is shown in the letter, the probability for a singlet valence bond of length $L$ to arise is of order $|v/u|^{2L}$, so it is sufficient to keep only the $L=1$ valence bonds. A singlet valence bond of length $L=1$ has a wave function (that is not normalized) $|b\rangle_{ij}=\sum_m(\psi^\dag_{i,m}\psi_{j,m}+\psi_{i,m}\psi^\dag_{j,m})\prod_k|s\rangle_k$. Define a valence-bond energy $V(\beta^2)=_{ij}\langle b|H_{I}(i\ \&\ j)|b\rangle_{ij}$, where $H_{I}(i\ \&\ j)$ is the interaction Hamiltonian on site $i$ and $j$ only. (To be clear, the energy of a singlet valence bond of length $L=1$ is $V(\beta^2)/_{ij}\langle b|b\rangle_{ij}$ instead of simply $V(\beta^2)$, since the wave function is not normalized.) After a careful calculation, this energy is shown to be
\begin{equation}
V(\beta^2)= \frac{2}{5}\left\{(2+\beta^2)\left[7G_{32}(1-\beta^2)+8G_{42}\beta^2\right]+(7+\beta^2)\left[2G_{12}(1-\beta^2)+3G_{22}\beta^2\right]\right\}\ ,
\end{equation}
where we have defined $G_{32}=E_{321}-E_{200}$, $G_{42}=(11E_{424}+E_{422})/12-E_{200}$, $G_{12}=E_{121}-E_{200}$, and $G_{22}=E_{222}-E_{200}$, in terms of the energies computed in Tab. \ref{I}. All of them are positive and of order $E_a$. We note that $V(\beta^2)$ is a quadratic function of $\beta^2$. Since the total number of bonds is $zN_S/2$, each site owns in average $z/2$ bonds. The interaction energy per site is then
\begin{equation}
\langle H_I\rangle/N_S\approx |u|^2\beta^2\Delta+\frac{1}{2}z|v|^2V(\beta^2)\ ,
\end{equation}
where the first term comes from the energy of the background singlet state $|s\rangle_i$.

In the estimation of the hopping energy we assume $t>0$ is real and positive, namely we do not consider any magnetic flux. This allows us to set both $\alpha$ and $\beta$ real on all sites (we have already done so in the above). By directly acting with $H_t$ onto the wave function $|SLC\rangle$, we can represent the hopping energy per site up to quadratic order $|v/u|^2$ after a rearrangement as
\begin{equation}
\begin{split}
&\langle H_t\rangle/N_S=-\frac{zt}{2}(u+u*)(v+v*)\sum_{m}\left({_i\langle} s|\psi_{i,m}\psi^\dag_{i,m}|s\rangle_i\right) \left({_j\langle} s|\psi^\dag_{j,m}\psi_{j,m}|s\rangle_j\right)\\
&\qquad-2\times \frac{zt}{2}|v|^2\sum_{m_1,m_2,m_3}\Big[\left({_i\langle} s|\psi_{i,m_1}\psi_{i,m_2}\psi^\dag_{i,m_3}|s\rangle_i\right) \left({_j\langle} s|\psi^\dag_{j,m_1}\psi^\dag_{j,m_2}\psi_{j,m_3}|s\rangle_j\right)+ (i\leftrightarrow j)^\dag\\
&\qquad+\left({_i\langle} s|\psi_{i,m_1}\psi^\dag_{i,m_2}\psi_{i,m_3}|s\rangle_i\right) \left({_j\langle} s|\psi^\dag_{j,m_1}\psi_{j,m_2}\psi^\dag_{j,m_3}|s\rangle_j\right)\Big]\ .
\end{split}
\end{equation}
Calculation of the first term is straightforward. To calculate the second term, one can use the Wigner-Eckart theorem in the group theory, which tells us that
\begin{equation}
{_i\langle} s|\psi_{i,m_1}\psi_{i,m_2}\psi^\dag_{i,m_3}|s\rangle_i=c\delta_{m_1+m_2,m_3}\sqrt{5}\left(\begin{array}{ccc}2&2&2\\m_1&m_2&-m_1-m_2\\ \end{array}\right)\ ,
\end{equation}
where we have used the Wigner-$3j$ symbol instead of the Clesch-Gordan coefficient. $c$ here is a coefficient. It is easy to find $c=2\sqrt{3}/5$ by calculating an example. Similar relations with the same coefficient $c=2\sqrt{3}/5$ hold for the other three-boson operator expectation values. The hopping energy is then calculated to be
\begin{equation}
\begin{split}
&\langle H_t\rangle/N_S=-\frac{zt}{2}(u+u^*)(v+v^*)5\cdot\frac{2\alpha^2+3\beta^2}{5}\cdot\frac{7\alpha^2+8\beta^2}{5}\\
&-zt|v|^2\alpha^2\beta^2\sum_{m_1,m_2}5\left(\begin{array}{ccc}2&2&2\\m_1&m_2&-m_1-m_2\\ \end{array}\right)^2\left[\left(\frac{2\sqrt{3}}{5}\right)^2+\left(\frac{2\sqrt{3}}{5}\right)^2+\left(\frac{2\sqrt{3}}{5}\right)^2\right]\\
&=-\frac{2}{5}zt\text{Re}(u)\text{Re}(v)(2+\beta^2)(7+\beta^2)-\frac{36}{5}zt|v|^2(1-\beta^2)\beta^2\ .
\end{split}
\end{equation}
Putting the two parts of energy together, we have
\begin{equation}
\mathcal{E}_G=|u|^2\beta^2\Delta-\frac{2}{5}zt\text{Re}(u)\text{Re}(v)\left(2+\beta^2\right)\left(7+\beta^2\right)
-\frac{36}{5}zt|v|^2\left(1-\beta^2\right)\beta^2+\frac{1}{2}z|v|^2V\left(\beta^2\right),
\end{equation}
where $0\le\beta^2\le1$ is imposed. Obviously, for the energy to be the lowest, both $u$ and $v$ should be real and positive, as is assumed then.

Now we optimize the variational energy, and derive the phase boundary of the SLC phase. By Eq. (\ref{nor}) in the above we can eliminate $u$ and write $\mathcal{E}_G$ as a function of $\beta^2$ and $v$. The minimum energy in the SLC phase is achieved when
\begin{equation}
\begin{split}
0&=\frac{1}{v}\frac{\partial \mathcal{E}_G}{\partial v}=-\frac{4}{5}(2+\beta^2)(7+\beta^2)\beta^2\Delta+zV(\beta^2)-\frac{72}{5}zt(1-\beta^2)\beta^2\\
&\qquad\qquad\qquad -zt\left[\frac{2}{5}\left(\frac{u}{v}\right)(2+\beta^2)(7+\beta^2)-\frac{4}{25}\left(\frac{v}{u}\right)(2+\beta^2)^2(7+\beta^2)^2\right]\ ,\\
0&=\frac{1}{u^2}\frac{\partial \mathcal{E}_G}{\partial (\beta^2)}=\Delta-\frac{2}{5}zt\left(\frac{v}{u}\right)(9+2\beta^2) +\frac{1}{5}z\left(\frac{v}{u}\right)^2\left[\frac{5}{2}\frac{\partial V(\beta^2)}{\partial (\beta^2)}-2(9\beta^2 +2\beta^4)\Delta-36t(1-2\beta^2)\right]\ .
\end{split}
\end{equation}
If the system is right on the SLC phase boundary, the above conditions should give exactly $\beta^2=0$ or $\beta^2=1$. By setting $\beta^2=0$, one obtains the phase boundary between SLC and the dimer MI, given by
\begin{equation}\label{b0}
\begin{split}
&\qquad\qquad\qquad\qquad \frac{28}{5}t\left(\frac{v}{u}\right)^2+(G_{32}+G_{12})\left(\frac{v}{u}\right)-t=0\ ,\\
&\Delta-\frac{18}{5}zt\left(\frac{v}{u}\right)+\frac{1}{5}z\left(\frac{v}{u}\right)^2(16G_{42}-7G_{32}+21G_{22}-12G_{12}) -\frac{36}{5}zt\left(\frac{v}{u}\right)^2=0\ .
\end{split}
\end{equation}
For $\beta^2=1$, the phase boundary between SLC and the trimer MI is given by
\begin{equation}\label{b1}
\begin{split}
&\qquad\qquad\qquad\qquad \frac{48}{5}t\left(\frac{v}{u}\right)^2+(G_{42}+G_{22}-2\Delta)\left(\frac{v}{u}\right)-t=0\ ,\\
&\Delta-\frac{22}{5}zt\left(\frac{v}{u}\right)+\frac{1}{5}z\left(\frac{v}{u}\right)^2(32G_{42}-21G_{32}+27G_{22}-16G_{12}-22\Delta) +\frac{36}{5}zt\left(\frac{v}{u}\right)^2=0\ .
\end{split}
\end{equation}
To find the explicit expression of the phase boundaries, one has to eliminate $(v/u)$ from Eqs. (\ref{b0}) and (\ref{b1}). In the elimination, one should ensure $(v/u)>0$ for the result to be physical. One can further verify that $0<|\beta^2|<1$ in between the two phase boundaries. In Fig. 2 in the main text, we have plotted explicitly the phase boundaries for $U_4=3U_2=30U_0$ and $z=4$. In particular in Fig. \ref{beta} below, we have calculated how the value of $|\beta|^2$ varies with respect to the chemical potential $\mu$ for $zt/U_0=5.5$. It can be explicitly seen that the system undergoes the phase transitions from dimer MI to SLC and then to trimer MI.

\begin{figure}
\includegraphics[width=3.2in]{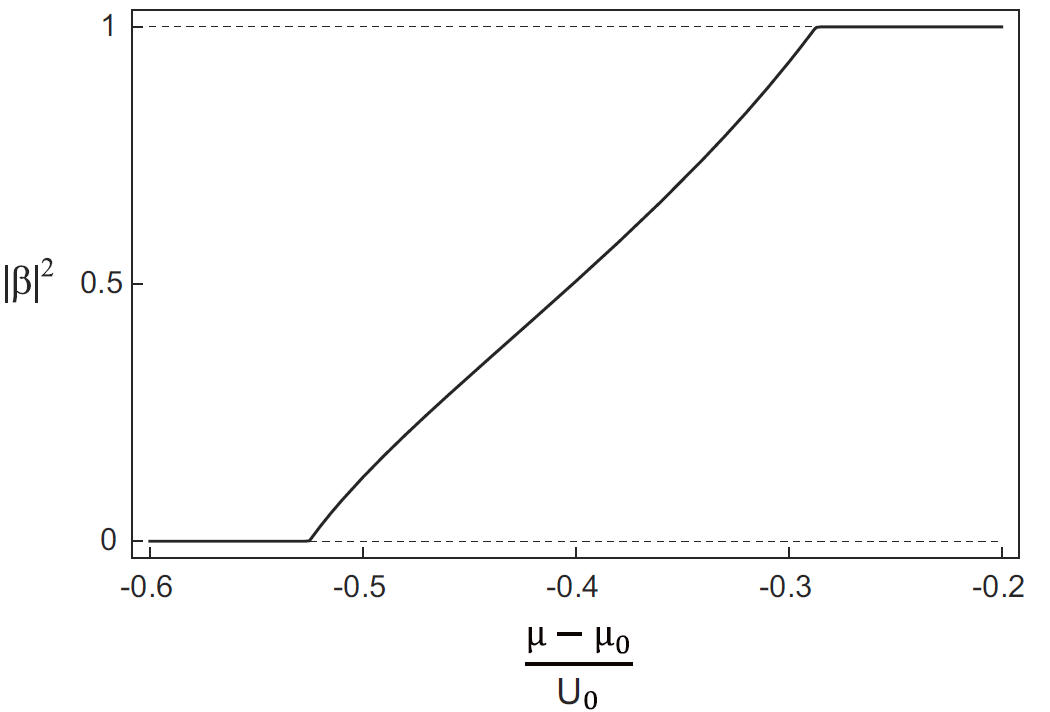}
\caption{The value of $|\beta|^2$ where the variational energy is minimized for $zt/U_0=5.5$. As one increase the chemical potential $\mu$, $|\beta|^2$ becomes nonzero at some point, gradually increases to $1$ and then stop. This indicates the phase transitions from dimer MI to SLC and then to trimer MI. \label{beta}}
\end{figure}

In the limit $t/E_a\ll1$, one finds in both cases $v/u\propto t/E_a$, so the phase boundaries have the limiting form:
\begin{equation}
\Delta+a_{\pm}\frac{zt^2}{E_a}=\pm b_\pm^2\frac{zt^3}{E_a^2}\ ,
\end{equation}
where $a_\pm$ and $b_{\pm}$ are coefficients depending on the interactions $U_{2J}$ only. As is defined before, $\Delta$ can be further replaced by $\mu_0-\mu$.

\subsection{Determination of the phase boundary of spinor BEC}

The phase boundary between Mott insulator and spinor BEC is obtained based on the Gutzwiller variational wave function method used in many previous studies. We first consider the transition from a dimer MI to a spinor BEC. The method proposes the following direct product wave function:
\begin{equation}
|\varphi\rangle=\prod_i\left(\sqrt{1-|\eta|^2-|\xi|^2}|2,0,0\rangle_i+\eta\sum_m\varphi_m|3,2,m\rangle_i +\xi^*\sum_m(-1)^m\varphi_m^*|1,2,-m\rangle_i\right)\ ,
\end{equation}
where $\varphi$ is a normalized spinor that characterizes the spinor BEC phase. The energy expectation of this wave function can be easily derived as:
\begin{equation}
\begin{split}
\mathcal{E}_{\text{spinor}}&=|\eta|^2E_{32}+|\xi|^2E_{12}-zt\left(1-|\eta|^2-|\xi|^2\right)\left|\sqrt{\frac{7}{5}}\eta+\sqrt{\frac{2}{5}}\xi\right|^2 \\
&=(\eta^*,\xi^*)\left(\begin{array}{cc}E_{32}-\frac{7}{5}zt&-\frac{\sqrt{14}}{5}zt\\-\frac{\sqrt{14}}{5}zt&E_{12}-\frac{2}{5}zt\\ \end{array}\right)\left(\begin{array}{c}\eta \\ \xi\end{array}\right) +\text{higher order terms}\ ,
\end{split}
\end{equation}
When the quadratic term $(\eta^*,\xi^*)M(\eta,\xi)^T$ becomes non-positive, where $M$ stands for the $2\times2$ matrix in the above, the system falls into a spinor BEC phase. So the phase boundary is simply determined by the condition $\det M=0$. Similarly we can obtain the phase boundary between the trimer MI and the the spinor BEC phase. The phase boundary between SLC and spinor BEC can then be obtained via an interpolation.

However, this method is only accurate to the first order of $t/E_a$. With higher order corrections, the phase boundary should be further modified. On the other hand, the calculation for the SLC phase in the previous section is done up to the third order of $t/E_a$, which may not match very well with the calculation here for spinor BEC. Therefore, we add a higher order correction to the spinor BEC phase boundary, so that the predicted triple point of the dimer MI, trimer MI and the spinor BEC calculated using this Gutzwiller method is located inside the SLC phase obtained in the last section. The resulting phase diagram is shown in Fig. 2. We further note that according to Ref. \cite{Ciobanu2000,Ueda2002,Barnett2006}, the spinor BEC phase in the case $U_4=3U_2=30U_0$ should be a spin $2$ nematic phase.

\subsection{Mapping between the norm of the SLC state and the loop gas}
The classical loop gas model has been used to study the RVB state in fermionic systems \cite{Sutherland1988}. Following the same idea, we can also construct the loop gas model in equivalent to the norm of the SLC wave function $|SLC\rangle$. The norm of $|SLC\rangle$ can then be calculated numerically with the loop gas model. Following Ref. \cite{Sutherland1988}, we briefly sketch the mapping here.

As a simplest mapping, we keep only the contributions of singlet resonating valence bonds of length $L=1$. The non-normalized wave function of one such valence bond is $|b\rangle_{ij}$ as is defined previously. Similarly, we can construct the wave function $|b_\mathcal{M}\rangle$ of an arbitrary valence bond ($L=1$) configuration $\mathcal{M}$,
\begin{equation}
|b_{\mathcal{M}}=\prod_{\langle ij\rangle\in\mathcal{M}}\left[\sum_m(\psi^\dag_{i,m}\psi_{j,m}+\psi_{i,m}\psi^\dag_{j,m})\right]\prod_{k}|s\rangle_k\ ,
\end{equation}
where a site can at most connect to one valence bond ($L=1$). In this approximation, one finds the wave function $|SLC\rangle$ the superposition of all kinds of such configurations
\begin{equation}
|SLC\rangle=u^{zN_S/2}\prod_i|s\rangle_i+\sum_{\mathcal{M}}v^{L_\mathcal{M}}u^{zN_S/2-L_\mathcal{M}}|b_\mathcal{M}\rangle\ ,
\end{equation}
where $L_\mathcal{M}$ is the total number of valence bonds in the configuration $\mathcal{M}$. We note that all the coefficients of superposition are real and positive, as is concluded previously.

To calculate the norm of $|SLC\rangle$, one has to calculate the overlap between two configurations $\langle b_{\mathcal{M}}|b_{\mathcal{M'}}\rangle$. Analogous to the argument in Ref. \cite{Sutherland1988}, the overlap is non-zero only if the overlap of $\mathcal{M}$ and $\mathcal{M'}$ consists only of closed loops of valence bonds. In the calculation, the loops of length $L_c=2$ (formed by a bond and itself) have a different contribution from those of loops of length $L_c>2$. Besides, a loop has a non-zero contribution only if $L_c$ is even (which is always satisfied in a square or cubic lattice), since no odd length loop can occur in the overlap of $\mathcal{M}$ and $\mathcal{M}'$. Concretely, a $L_c=2$ loop's contribution is given by
\begin{equation}
v^2u^{zN_S/2-2}{_{ij}}\langle b| b\rangle_{ij}=u^{zN_S/2}\left(\frac{v}{u}\right)^2\cdot\frac{2}{5}(2+\beta^2)(7+\beta^2)=10u^{zN_S/2}e^{-2\epsilon_b}\ ,
\end{equation}
while that of a $L_c>2$ loop $C$ is given by
\begin{equation}
2v^{L_c}u^{zN_S/2-L_c}\prod_{i=1\in C}^{L_c}{_{i-1,i}}\langle b| b\rangle_{i,i+1}=2u^{zN_S/2}\left(\frac{v}{u}\right)^{L_c}\cdot 20\left[\frac{(2+\beta^2)(7+\beta^2)}{25}\right]^{L_c/2}=20u^{zN_S/2}e^{-L_c\epsilon_b}\ ,
\end{equation}
where we have defined an ``energy" $\epsilon_b$, and the additional factor $2$ for $L_c>2$ loops comes from the fact that the (alternating) bonds can come from either $\mathcal{M}$ or $\mathcal{M}'$ \cite{Sutherland1988}. Therefore, if we use $\mathcal{M}_C$ to denote a configuration of non-intersecting loops with even lengths, where there are $P_{L_c}(\mathcal{M}_C)$ number of loops with length $L_c$, the norm of wave function $|SLC\rangle$ is given by
\begin{equation}
\langle SLC|SLC\rangle=u^{zN_S/2}\sum_{\mathcal{M}_C}e^{-\epsilon_bL(\mathcal{M}_C)+P(\mathcal{M}_C)\ln20-P_2(\mathcal{M}_C)\ln2}\ ,
\end{equation}
where $P(\mathcal{M}_C)=\sum_{L_c}P_{L_c}$ is the total number of loops, and $L(\mathcal{M}_C)=\sum_{L_c}L_cP_{L_c}$ is the total length of all the loops. This expression can be viewed as the partition function of a classical loop gas model, where the energy of a loop is proportional to its length, and the chemical potential of $L_c=2$ loops differs from that of $L_c>2$ loops. The norm can therefore be calculated using a Monte Carlo method.

In principle, the calculation of the variational energy can also be embedded in the loop gas model \cite{Sutherland1988}, which shall not discuss the details here. We note that different from the loop gas model for RVB state, the loop gas model here does not require each site to be connected to a loop. The loop configurations here can then consist of very few loops.

In this simplest approximation, the loop gas model contains only non-intersecting loops, namely has an infinite contact repulsion between the loops. If one keeps singlet valence bonds with length $L>1$, one could obtain a model where the loops can intersect with each other with an interaction energy. This is more complicated and beyond the discussion here.

\end{widetext}

\end{document}